\documentclass[showpacs,amssymb,preprint,preprintnumbers,nofootinbib,superscriptaddress]{revtex4-2}
\usepackage{amsmath}
\usepackage{graphicx}
\graphicspath{{Figures/}}
\usepackage{latexsym}
\usepackage{amsfonts}
\usepackage{url}
\usepackage{hyperref}
\usepackage{bm}
\usepackage{textcomp}
\usepackage{color}
\usepackage{bbm}
\usepackage{slashed}
\usepackage{caption}
\usepackage{epstopdf}
\usepackage{subcaption}
\usepackage{placeins}
\usepackage{color}
\usepackage{cancel} 
\usepackage[normalem]{ulem}
\usepackage{tcolorbox}
\usepackage{soul}

\newcommand{\Ren}{R\'enyi }
\newcommand{\R}{\text{R}}
\newcommand{\dd}{\text{d}}
\newcommand{\ba}{\begin{eqnarray}}
\newcommand{\ea}{\end{eqnarray}}
\newcommand{\la}{\lambda}
\newcommand{\lt}{\left}
\newcommand{\rt}{\right}
\newcommand{\pa}{\partial}

\newcommand{\rde}{\text{RHDE}}
\newcommand{\La}{\Lambda}

\begin{document}

\title{\Ren Holographic Dark Energy}

\author{Ratchaphat Nakarachinda \footnote{Email: tahpahctar\_net@hotmail.com}}
\affiliation{The Institute for Fundamental Study (IF), Naresuan University,\\ 99 Moo 9, Tah Poe, Mueang Phitsanulok, Phitsanulok, 65000, Thailand}
\affiliation{Department of Physics, School of Science, King Mongkut's Institute of Technology Ladkrabang, 1 Chalong Krung 1 Alley, Lat Krabang, Bangkok 10520, Thailand}	
		
\author{Chakrit Pongkitivanichkul \footnote{Email: chakpo@kku.ac.th}}
\affiliation{Khon Kaen Particle Physics and Cosmology Theory Group (KKPaCT), Department of Physics, Faculty of Science, Khon Kaen University, 123 Mitraphap Rd., Khon Kaen, 40002, Thailand}

\author{\\Daris Samart \footnote{Email: darisa@kku.ac.th}}
\affiliation{Khon Kaen Particle Physics and Cosmology Theory Group (KKPaCT), Department of Physics, Faculty of Science, Khon Kaen University, 123 Mitraphap Rd., Khon Kaen, 40002, Thailand}

\author{Lunchakorn Tannukij \footnote{Email: l\_tannukij@hotmail.com}}
\affiliation{Department of Physics, School of Science, King Mongkut's Institute of Technology Ladkrabang, 1 Chalong Krung 1 Alley, Lat Krabang, Bangkok 10520, Thailand}

\author{Pitayuth Wongjun \footnote{Email: pitbaa@gmail.com}}
\affiliation{The Institute for Fundamental Study (IF), Naresuan University,\\ 99 Moo 9, Tah Poe, Mueang Phitsanulok, Phitsanulok, 65000, Thailand}


\begin{abstract}

In this work, the holographic dark energy model is constructed by using the non-extensive nature of the Schwarzschild black hole via the \Ren entropy.
Due to the non-extensivity, the black hole can be stable under the process of fixing the non-extensive parameter.
A change undergoing such a process would then motivate us to define the energy density of the \Ren holographic dark energy (RHDE).
As a result, the RHDE with choosing the characteristic length scale as the Hubble radius provides the late-time expansion without the issue of causality.
Remarkably, the proposed dark energy model contains the non-extensive length scale parameter additional to the standard $\La$CDM model.
The cosmic evolution can be characterized by comparing the size of the Universe to this length scale.
Moreover, the preferable value of the non-extensive length scale is determined by fitting the model to recent observations.
The results of this work would shed light on the interplay between the thermodynamic description of the black hole with non-extensivity and the classical gravity description of the evolution of the Universe.

\end{abstract}
\maketitle{}


\newpage

\section{Introduction}\label{sec: intro}

One of the significant theoretical challenges in modern cosmology is an attempt to explain the accelerated expansion of the Universe nowadays \cite{SupernovaSearchTeam:1998fmf,SupernovaCosmologyProject:1998vns}. It is possible to categorize such theoretical models into two classes, modified gravity theories and dark energy, adding exotic contributions to the Einstein field equation. The simple and promising candidate for dark energy is cosmological constant which is well-known as $\Lambda$CDM. Since its energy density of the cosmological constant is constant throughout the whole evolution of the Universe, it is consequently interpreted as the vacuum energy \cite{Condon:2018eqx,Peebles:2002gy}. However, the value of the energy density calculated from standard quantum field theory has a discrepancy of about 120 orders in magnitude with the observed value \cite{Weinberg:1988cp}. This leads to the fine-tuning problem in which the value of the energy density in the past must be tremendously fine-tuned. Moreover, this leads to the puzzle of why the contributions from dark energy and non-relativistic matter are comparable nowadays, namely coincidence problem \cite{Zlatev:1998tr}. As a result, numerous dynamical models of dark energy have been proposed \cite{Copeland:2006wr,Shapiro:2009dh}. In this article, we are interested in the dark energy model in which the root idea is inspired by a fundamental principle called the holographic principle. Therefore, such a dark energy model is commonly known as ``holographic dark energy'' (HDE).

The HDE is rooted from the holographic principle which generally refers
to the duality between theories of the bulk and its boundary \cite{Susskind:1994vu,Maldacena:1997re,Bousso:2002ju}. By using this notion with dimensional reduction, it is possible to prevent one from over-counting the degrees of freedom for the black hole entropy \cite{tHooft:1993dmi}. This provides a hint to solve the cosmological constant since it suggests that one may over-count the degrees of freedom by using local quantum field theory. Based on this notion, it may be a link between the effective field theory
which connects the energy density and length scale through
a saturation entropy. In other words, the ordinary matter with saturated energy should not form a black hole, i.e., $E_\text{HDE}\sim E_{\text{BH}}=L/(2G)$.
According to such a definition, there exists a relationship between the IR scale corresponding to the characteristic length scale of the dark energy $L$ and the UV scale corresponding to the energy scale of the black hole or the Newtonian gravitational constant $G$. Note that such a relation is also obtained when we consider the Bekenstein bound \cite{Bekenstein:1973ur, Hawking:1975vcx, Bardeen:1973gs}, but it is expressed by considering the entropy \cite{Cohen:1998zx}.
As a result, the energy density of HDE is constructed as follows 
\ba
    \rho_\text{HDE}=\frac{3b^2}{8\pi GL^2},\label{rho orig}
\ea
where $b^2$ is a dimensionless constant. 
It is seen that the characteristic length scale is arbitrary.

In describing the dynamics of the Universe, the IR length scale $L$ should be interpreted as some cosmological length.
One of the possibilities is the Hubble radius.
Unfortunately, the model of HDE with Hubble radius as the IR length scale cannot drive the accelerated expansion \cite{Hsu:2004ri}.
In fact, the evolution of this HDE is the same as the dominant species in the Universe.
For example, in the matter-dominated epoch, the energy density is proportional to $a^{-3}$ where $a$ is the scale factor. 
This HDE is therefore diluted as time grows.
After that the successful model is proposed by treating the IR length scale as the future particle horizon \cite{Li:2004rb}.
The cosmological implications of this model are investigated in various aspects \cite{Wang:2016och}.
For example, many types of interaction between HDE and dark matter and their observational constraints are studied \cite{Karwan:2008ig, Ma:2009uw, Zhang:2012uu}. 
Even though this model can drive the accelerated expansion of the Universe, it suffers from a problem of causality \cite{Colgain:2021beg}.
The information in the future is required to describe the dynamics at the present time.
However, there is an attempt to overcome this problem by constructing an action consistent with the Friedmann equation. As a consequence of this consideration, it may be possible to interpret the use of the future particle horizon as the characteristic length scale in terms of the initial conditions at present \cite{Li:2012xf, Li:2013}.

Other models with different IR length scales are proposed \cite{Sheykhi:2009zv,Huang:2012nz,Huang:2012xma,Nojiri:2019skr,Nojiri:2021iko}.
In addition, HDE in modified gravity theories are also numerously active \cite{Nojiri:2005pu,Banerjee:2007zd,Bisabr:2008gu,Kritpetch:2020vea,Saridakis:2007ns,Farajollahi:2014hzp}.
Based on the holographic principle, it is feasible to construct more types of dark energy models. 
Using time scale as the IR cutoff, the agegraphic dark energy concept has been put out \cite{Cai:2007us,Wei:2007ty}.
Since the Ricci scalar $R$ does in fact have a dimension of $L^{-2}$, one can define the dark energy based on the corresponding length scale,
referred to as the Ricci dark energy 
\cite{Gao:2007ep,Granda:2008dk}.
The perturbation analysis of the Ricci model is also investigated \cite{Karwan:2011sh}.
Since the notion of HDE roots from the black hole entropy and there are various investigations on the generalized entropies, the HDE model with generalized entropies have been investigated  \cite{Komatsu:2016vof,SayahianJahromi:2018irq,Dabrowski:2020atl,Saridakis:2020zol,Anagnostopoulos:2020ctz}.

According to the thermodynamic properties of the cosmological horizon \cite{Jacobson:1995ab,Cai:2005ra,Akbar:2006kj}, there is a worthwhile proposal to extend the form of the energy density of HDE by involving the first law of thermodynamics at the horizon \cite{Moradpour:2018ivi}. Specifically, the energy density of the HDE can be written in derivative form as $\rho_\text{HDE} = T \dd S/\dd V$. In this form, it provides a consistent way to deal with various kinds of generalized entropy. Therefore, there have been intensive investigations using this form of the energy density \cite{Sheykhi:2021fwh,Lymperis:2018iuz,Nojiri:2022aof,Nojiri:2021jxf,Nojiri:2022dkr}. Note that there is a suggestion that the Friedmann equations are also modified when the generalized entropy is taken into account \cite{Golanbari:2020coz}. Recently, rigorous investigation on this issue has been clarified by using the continuity equation \cite{Manoharan:2022qll}. It was shown that this form of the energy density is consistent with the standard HDE when we consider the Friedmann equations or entropies that follow exponent stretched area law such as Barrow entropy, Tsallis entropy, and Tsallis-Cirto entropy. For the R\'enyi entropy, the energy density will be explored and it will be examined whether cosmic accelerated expansion is achieved.

It is important to note that the mentioned investigations were performed according to horizon thermodynamics-gravity prescription by Jacobson \cite{Jacobson:1995ab} which suggests that any modification to the entropy leads to the modified gravitational field equations such as Friedmann equation. Therefore, in principle, gravitational field equations must be modified by introducing any modified entropies. However, if the gravitational field equations are supposed to be modified, it does not guarantee that the black hole can form and the Schwarzschild black hole may not exist. As a result, the original proposal by Cohen et al. may not be satisfied since the size of the black hole in the modified gravity is not determined. In order to keep the proposal by Cohen et al. \cite{Cohen:1998zx} satisfying as well as the modified entropy applying, one can consider the thermodynamics as the properties of the black hole. In this regime, the thermodynamics with generalized entropies can be obtained by using Euler’s homogeneous function theorem without modified gravitational equations \cite{Nakarachinda:2022gsb}. In the present work, we adopt the first law of thermodynamics based on the properties of the black hole instead of using one in thermodynamics at the apparent horizon as commonly found in the literature. Note that, in our approach, it can be applied to other kinds of black holes and then the energy density of HDE can be modified without modified gravitational equations. 
Actually, by using the first law of thermodynamics of the AdS black hole, the accelerated expansion with proper evolution of the Universe can be obtained \cite{Nakarachinda:2022mlz}. 
As a result, the form of energy density in Eq.~\eqref{rho orig} can be obtained from the first law of black hole thermodynamics,
\ba
    \rho_\text{HDE}
    \sim\frac{\dd M_\text{BH}}{\dd V}
    =T_\text{G}\frac{\dd S_\text{G}}{\dd V},
\ea
where $T_\text{G} = \partial M_\text{BH} /\partial S_\text{G}$ and $S_\text{G}$ are the conjugate variable for the entropy which is interpreted as temperature and the generalized  entropy, respectively.
$V = \partial M_\text{BH} /\partial P$ is a conjugate variable for the pressure $P$ simply interpreted as the volume of the Universe. Note that by keeping the temperature defined in this way, the Legendre structure is preserved then thermodynamic quantities are suitably defined.

From the black hole thermodynamics viewpoint, the thermal system associated with the black hole is a non-extensive system since the entropy of the black hole is adopted as the horizon area \cite{Hawking:1974rv,Bekenstein:1980jp}. Therefore, it is worthwhile to formulate the entropy of the black hole as non-extensive entropy. Accordingly, there are several investigations on the non-extensive nature of the black holes \cite{Nojiri:2021czz,Nojiri:2022sfd,Promsiri:2020jga,Promsiri:2021hhv,Tannukij:2020njz,Nakarachinda:2021jxd,Hirunsirisawat:2022fsb,Nakarachinda:2022gsb,Liu:2022snq,Majhi:2017zao,Luciano:2021mto,Abreu:2022pil,Cimdiker:2022ics,Cimidiker:2023kle,Sriling:2021lpr,Chunaksorn:2022whl,Jawad:2022lww,Luciano:2023fyr,Luciano:2023bai,Ghaffari:2023vcw}. One of the useful choices of the non-extensive entropy is the Tsallis entropy \cite{Tsallis:1987eu,Tsallis:2009zex}. However, by using Tsallis statistics, it is not easy to define a proper temperature which is the state variable based on the notion of maximum entropy. In fact, the composition rule for the Tsallis entropy does not provide the definition in such a way.  
The treatment is, fortunately, proposed by dealing with extensive thermodynamics in which the entropic function characterizing the system is the formal logarithm map of the Tsallis entropy \cite{Biro:2011}.
In an interesting coincidence, the form of the mapped entropic function is equivalent to the R\'enyi entropy \cite{Renyi1959}. 
A temperature compatible with the zeroth law is achieved by defining the so-called \Ren temperature as the proper temperature of black holes \cite{Biro:2013cra,Czinner:2015eyk,Czinner:2017bwc}.
Note that this temperature is identical to that defined from the analogy of Tsallis thermodynamics to the conventional extensive ones \cite{Jizba:2023ygi}.
In this sense, other thermodynamic state variables and thermodynamic laws can be defined consistently \cite{Nakarachinda:2022gsb,Cimidiker:2023kle}. 
It is a worthy result of the R\'enyi entropy as the black hole entropy. 
In this work, we utilize the R\'enyi entropy along with the first law of thermodynamics and examine how the accelerated expansion of the Universe can be achieved in terms of HDE.

For the black hole thermodynamics with R\'enyi entropy, the non-extensive parameter $\lambda$ can be promoted as the thermodynamic variable. This variable can be adopted as a kind of effective pressure of the system while its conjugate variable corresponds to the thermodynamics volume. To obtain a consistent energy density in terms of HDE, we consider the thermodynamics process at which the pressure is held fixed. Therefore, the non-extensive parameter is a model parameter in the cosmological aspect. With this setup, the model is similar to the HDE constructed from AdS black hole (AdS-HDE) \cite{Nakarachinda:2022mlz} while the non-extensive parameter can play the same role as the cosmological constant. As a result, we found that it is possible to obtain the accelerated expansion of the Universe similar to one in AdS-HDE. However, the interpretation is significantly different since the expansion of the Universe is due to the non-extensive nature of the black hole. By comparing the results to the observation, we found that the non-extensive length scale is preferable as $L_\lambda = \sqrt{G/(\pi \lambda)} \approx 0.07 H^{-1}_0$ where $H_0$ is the current observed value of the Hubble parameter. 
This may shed light on the interplay between the non-extensive nature of the black hole in the description of thermodynamics and the evolution of the Universe in the description of classical gravity.

Following is a description of how this work is organized.
Sec.~\ref{sec: bh} discusses the effect of nonextensivity on black hole thermodynamics by analyzing it via \Ren entropy.
The form of the energy density constructed from the black hole with \Ren statistics is defined in Sec.~\ref{sec: hde}.
The dynamics of the Universe with such dark energy are investigated in Sec.~\ref{sec: cosmo} and the model parameter is constrained by observational data in Sec.~\ref{sec: obs}.
Finally, in Sec.~\ref{sec: conclu}, the key results as well as the interesting remarks are discussed and summarized.


\section{Schwarzschild black hole with \Ren entropy}\label{sec: bh}

In this work, HDE motivated from \Ren thermodynamics of the Schwarzschild black hole is proposed to drive the late-time acceleration of the universe.  To formulate the corresponding energy density for the proposed HDE, the \Ren thermodynamics of Schwarzschild black hole is briefly reviewed in this section. 

A Schwarzschild black hole is the hole whose geometry is described, in terms of the Schwarzschild coordinates $(r,\theta,\phi)$, by the following Schwarzschild metric.
\begin{align}
\dd s^2 = -\left(1-\frac{2GM}{r}\right)\dd t^2+\left(1-\frac{2GM}{r}\right)^{-1}\dd r^2+r^2\dd\Omega^2,
\end{align}
where $M$ is the mass of the black hole which can be expressed in terms of the Schwarzschild radius, $r_h$, as 
\ba
    M=\frac{r_h}{2G}.\label{Sch mass}
\ea
The Hawking temperature, $T_\text{H}$, of such black hole and its Gibbs-Boltzmann entropy, $S_\text{BH}$, can be expressed in terms of $r_h$ as
\begin{align}
T_\text{H}=\frac{1}{4\pi r_h},\qquad S_\text{BH} = \frac{A}{4G}=\frac{\pi r_h^2}{G},
\end{align}
where $A$ is the area of the 2-sphere specified by the Schwarzschild radius. However, $S_\text{BH}$ is non-extensive since it does not scale with the size (3-volume) of the system. The non-extensivity can be incorporated into the consideration by assuming that the entropy of black holes may obey the following Tsallis composition rule \cite{Tsallis:1987eu}
\begin{align}
S_\text{tot} = S_1 + S_2 +\lambda S_1 S_2,
\end{align}
where $S_1$ and $S_2$ represent entropies of subsystems 1 and 2, $S_\text{tot}$ is the total entropy of the system whose components are the previously mentioned subsystems, and $\lambda$ denotes the non-extensivity of the system. Due to the ambiguity of realizing empirical temperature through the zeroth law of thermodynamics, one may transform the entropies associated with the Tsallis composition rule into the following \Ren entropy.
\begin{align}
S_\R = \frac{1}{\lambda} \ln{\left(1+\lambda S_\text{BH}\right)},
\label{SR in SBH}
\end{align}
which recovers the Gibbs-Boltzmann (GB) entropy when the non-extensive parameter $\lambda$ approaches zero.
The non-extensive parameter $\la$ plays both roles of the mapping parameter from non-additive Tsallis entropy to the additive \Ren one and of characterization of how much the \Ren entropy deviates from the standard GB entropy.
In addition, this parameter is bounded according to the statistical axiom of the maximized entropy as $\la<1$ in the unit of $k_\text{B}=1$ \cite{Renyi1959}.
It is clear from this bound that the entropic function behaves as a concave function.
According to the dS to AdS phase transition in five-dimensional spacetime with the Gauss-Bonnet gravity, however, there exists a negative value of $\la$ that produces a positive \Ren temperature \cite{Samart:2020klx}.
Note that the logarithmic function in Eq.~\eqref{SR in SBH} is always well-behaved when $\la$ is positive.
As a result, it is prudent to stick with our consideration for the range of the non-extensive parameter of $0\leq\la<1$.
In addition, the aforementioned range satisfies the strong condition of concavity in the context of quantum entanglement \cite{Horodecki:2009zz}.

\subsection{Thermodynamic phase space}

Since the non-extensivity $\lambda$ is introduced in the context of \Ren entropy, its thermodynamic phase space can be generally extended from the one in the context of GB entropy.
To construct the thermodynamic phase space due to \Ren entropy, the black hole's geometrical quantities must be expressed in the appropriate form \cite{Nakarachinda:2022gsb}.
It is found that the mass of the black hole in Eq.~\eqref{Sch mass} can be written in terms of a homogeneous function of degree $1/2$ of the \Ren entropy and the non-extensive parameter as follows:
\ba
	M(S_\R,\la^{-1})&=&\sqrt{\frac{e^{\la S_\R}-1}{4\pi G\la}}.
\ea
Applying Euler's theorem to this function, the Smarr formula and the first law are given by 
\ba
	M&=&2T_\R S_\R-2\Phi\la,\label{Smarr old}\\
	\dd M&=&T_\R\dd S_\R+\Phi\dd \la\label{1st old}.
\ea
Here, $T_\R$ and $\Phi$ are conjugate variables of $S_\R$ and $\la$, respectively.
They are defined as
\ba
	T_\R&=&\lt(\frac{\pa M}{\pa S_\R}\rt)_\la
	\,=\,\frac{1}{4\pi r_h}\lt(1+\la\frac{\pi r_h^2}{G}\rt),\\
	\Phi&=&\lt(\frac{\pa M}{\pa\la}\rt)_{S_\R}
	\,=\,\frac{1}{4\pi r_h\la^2}\lt[\lt(1+\la\frac{\pi r_h^2}{G}\rt)\ln\lt(1+\la\frac{\pi r_h^2}{G}\rt)-\la\frac{\pi r_h^2}{G}\rt].
\ea
Interestingly, the conjugate variable $\Phi$ can be approximated for small $\la$ as follows:
\ba
	\Phi&\approx&\frac{\pi r_h^3}{8G^2}-\la\frac{\pi^2r_h^5}{24G^3}+\la^2\frac{\pi^3r_h^7}{48G^4}+\hdots
\ea 
It is obvious that the leading term is proportional to the cubic of the horizon radius. 
This conjugate variable can be interpreted as a volume including the non-extensive effect.
Therefore, one, respectively, defines thermodynamic volume and pressure as
\ba
	V_\R=\frac{32G^2}{3}\Phi,
 \qquad
	P_\R=\frac{3}{32G^2}\la.\label{V P Ren}
\ea 
The factors in defining the volume and pressure are introduced in order to obtain that the leading order of $V_\R$ becomes $4\pi r_h^3/3$.  
Note that the ``pressure'' is interpreted via dimensional analysis which is suggested that the non-extensive parameter is of the same dimension as that of the thermodynamic pressure.
However, this pressure should not be mistaken for the thermodynamic pressure but should be considered in the same way as the cosmological constant is (considered similarly as pressure) in the context of the thermodynamics of black holes in the dS/AdS background (see Ref.~\cite{Kubiznak:2016qmn} and references therein). 
As a result, the thermodynamic relations~\eqref{Smarr old} and \eqref{1st old} are rewritten as
\ba
	M&=&2T_\R S_\R-2V_\R P_\R,\label{Smarr}\\
	\dd M&=&T_\R\dd S_\R+V_\R\dd P_\R\label{1st}.
\ea
According to the first law~\eqref{1st}, the mass of the black hole plays the role of the chemical enthalpy.
In addition, interpreting the non-extensive parameter as some sort of pressure has its own merit in the context of the energy in black hole creation. 
Since the mass of the black hole in this scenario is considered as the thermodynamic enthalpy, which is defined as $M=E+P_\R V_\R$, where $E$ is the usual thermodynamic internal energy, the non-extensive effect of the black hole may affect the energy required to form a black hole through $P_\R V_\R$ term where $P_\R$ arises from the non-extensive parameter $\la$. 
Following this reasoning, one may argue that the existence of the non-extensive nature of a black hole contributes to the creation of a black hole of mass $M$ as a work done by $P_\R V_\R$, in addition to the energy $E$.


\subsection{Stabilities of black hole}

\begin{figure}[!ht]
	\includegraphics[scale=0.27]{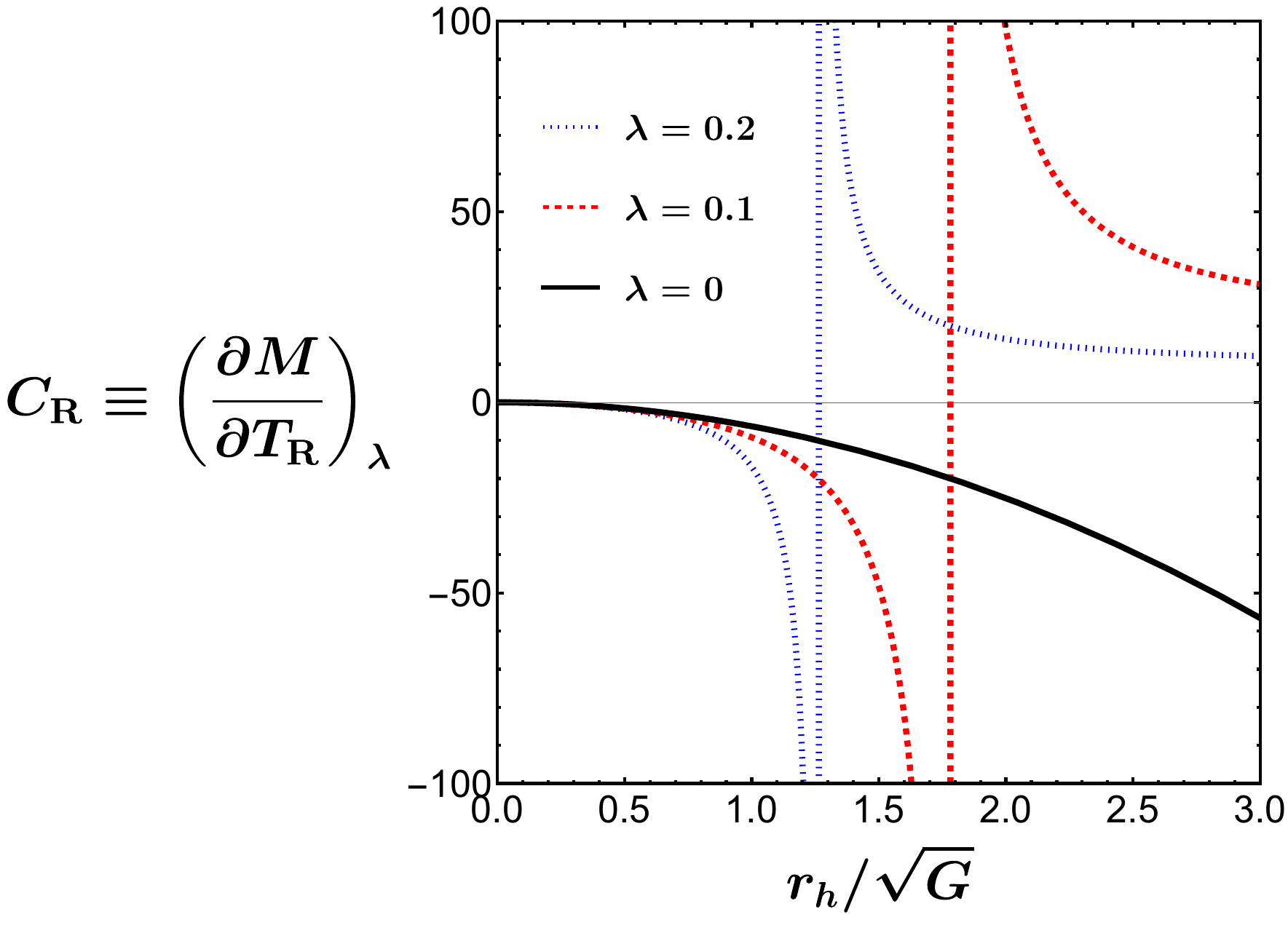}
	\hspace{1cm}
	\includegraphics[scale=0.27]{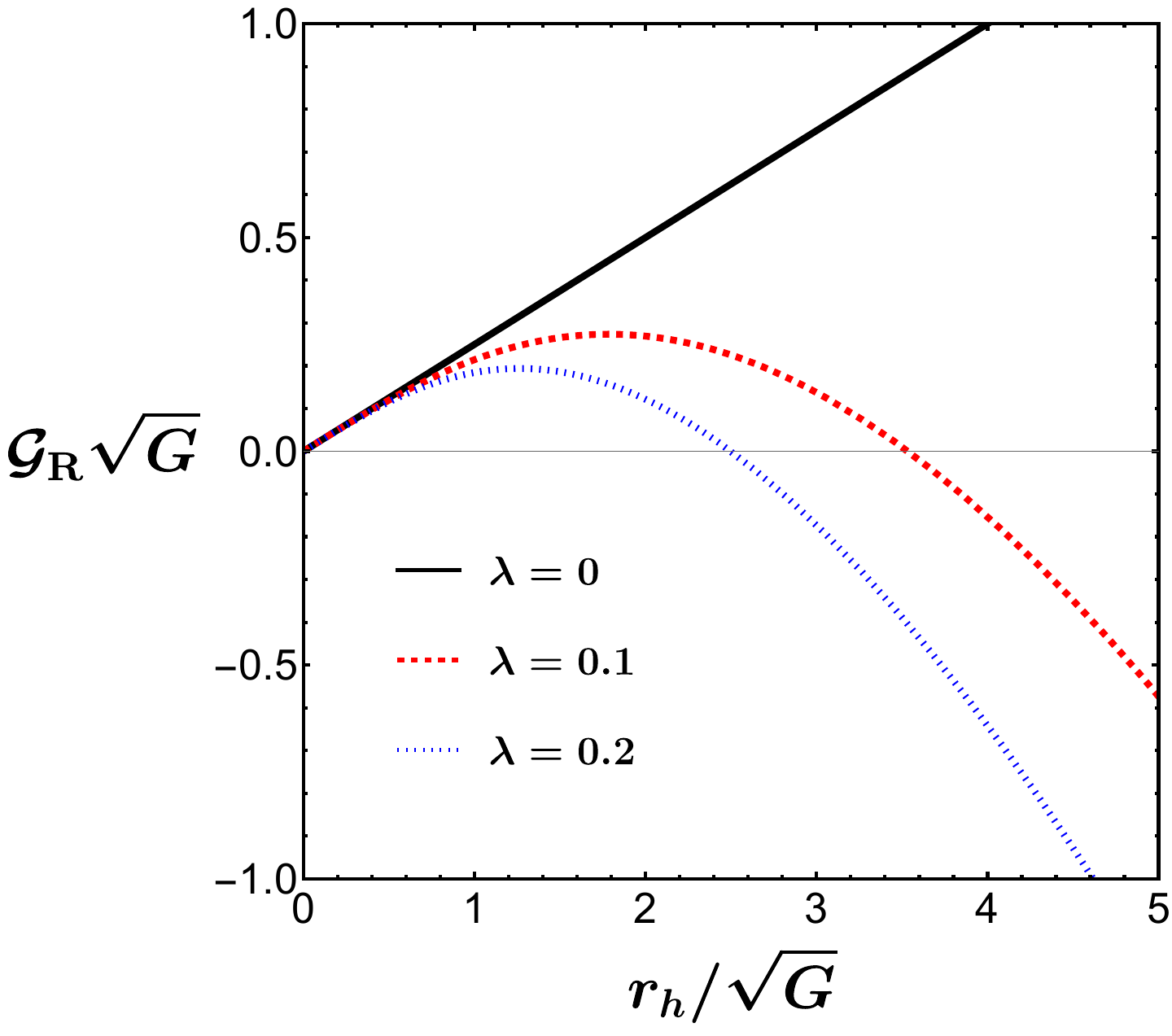}
\caption{Heat capacity (left) and Gibbs free energy (right) of Schwarzschild black hole with various values of non-extensive parameter $\la$}\label{fig:stab Sch}
\end{figure}

The thermodynamic stabilities of the black hole can be categorized into two cases. 
The first one is called local stability.
Such stability is characterized by the positiveness of the heat capacity, $C=\dd E/\dd T$. 
If the black hole has a negative heat capacity, it is locally unstable.
Imagine that the black hole is surrounded by an environment with a lower temperature. 
Despite trying to radiate its energy, it gets hotter. 
This is equivalent to the evaporation of the black hole, and it eventually disappears. 
On the other hand, if it is surrounded by an environment with a higher temperature, the black hole will absorb energy from the outside. 
Its temperature, however, gets progressively lower. 
In such a situation, the black hole is moving further and further away from equilibrium with the surrounding environment. 
Hence, it cannot be stable.
For an isobaric process, the heat capacity is given by
\ba
	C_{P_\R}&=&\lt(\frac{\pa M}{\pa T_\R}\rt)_{P_\R}
	\,=\,T_\R\lt(\frac{\pa S_\R}{\pa T_\R}\rt)_\la
	\,=\,-\frac{2\pi r_h^2}{G\lt(1-\la\frac{\pi r_h^2}{G}\rt)}.\label{C Ren}
\ea
It is seen that, without the non-extensive effect (i.e., $\la\to0$), the heat capacity of the black hole is always negative.
For $\la$ being positive, the heat capacity can be positive.
Moreover, the local stability of the black hole requires only positive (not negative) values of $\la$.
These results are illustrated in the left panel of Fig.~\ref{fig:stab Sch}. 
The black hole in the GB limit (black solid line) is always locally unstable while a large black hole described by \Ren entropy (e.g., those with blue dotted and red dashed lines) can be locally stable.
From Eq.~\eqref{C Ren}, the locally stable black hole must have a sufficiently large radius satisfied a condition:
\ba
    r_h>\sqrt{\frac{G}{\pi\la}}.\label{loc stab cond}
\ea

The second type of stability is called global stability.
This stability identifies which phase of all possibility preferably exists in nature.
In particular, the preferable phase must have lower free energy, e.g., Gibbs free energy, than others.
We are interested in whether the black hole is thermodynamically formed compared to the no-black hole phase (i.e., the free energy is set to be zero).
Consequently, the globally stable black hole is characterized as a phase with negative free energy. 
For our consideration, the Gibbs free energy is defined by
\ba
	\mathcal{G}_\R
	&=&M-T_\R S_\R
	\,=\,\frac{r_h}{2G}-\frac{1}{4\pi r_h\la}\lt(1+\la\frac{\pi r_h^2}{G}\rt)\ln\lt(1+\la\frac{\pi r_h^2}{G}\rt).
\ea
The effect of the non-extensivity appears only in the second term. 
It is found that this term (i.e., non-zero $\la$) can make $\mathcal{G}_\R$ negative for the sufficiently large black hole phase 
\ba
    r_h>\sqrt{\lt[-1+\exp\lt\{2+\text{ProductLog}\lt(-\frac{2}{e^2}\rt)\rt\}\rt]\frac{G}{\pi\la}}
    \approx1.98\sqrt{\frac{G}{\pi\la}},\label{glob stab cond}
\ea
where $\text{ProductLog}(y)$ is the solution for $z$ in  $y=ze^z$. 
It can be seen in the right panel of Fig.~\ref{fig:stab Sch}.
Again, the black hole described by the GB entropy is always globally unstable.

\begin{figure}[!ht]
	\includegraphics[scale=0.27]{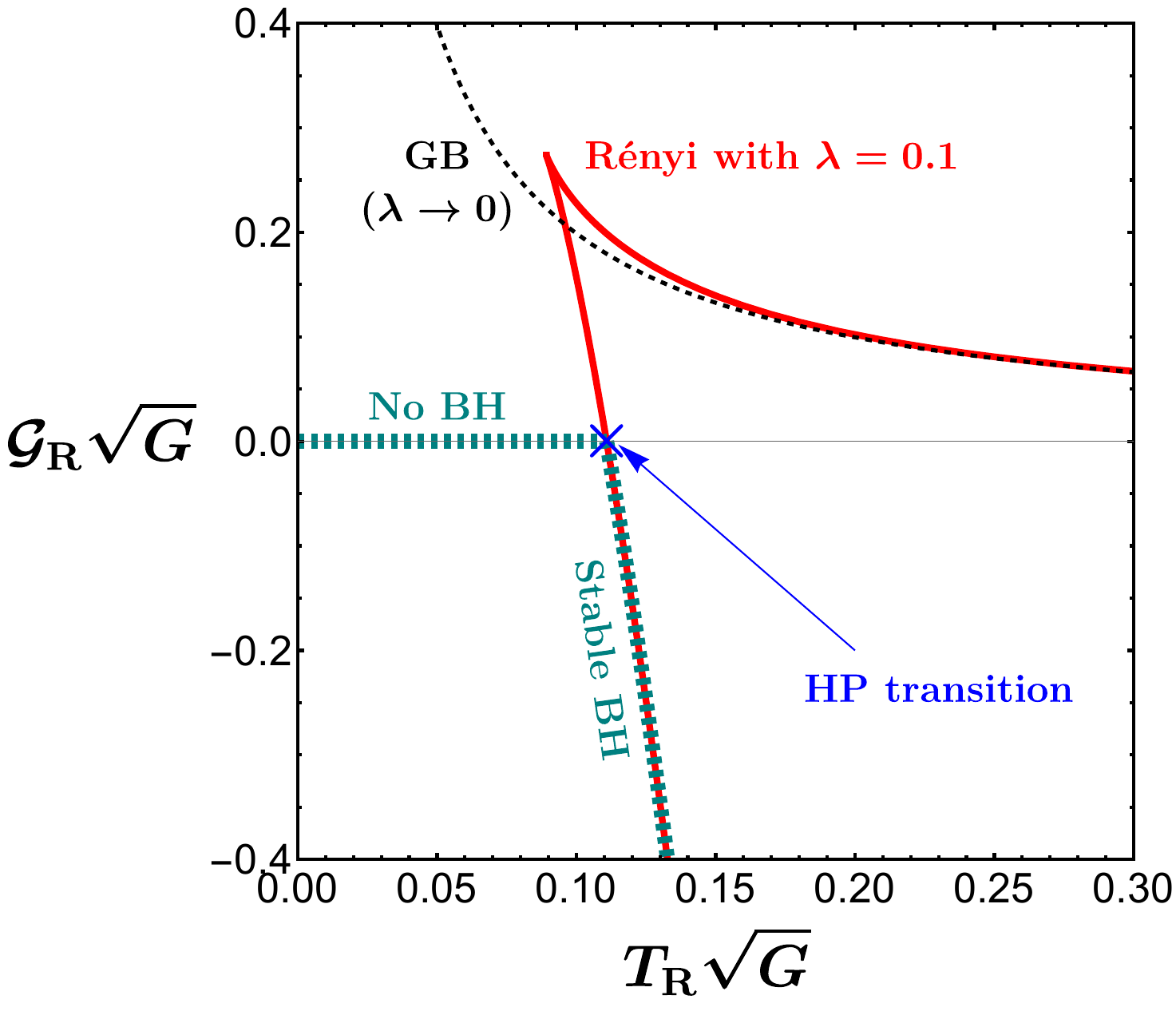}
\caption{Diagram of the Gibbs free energy versus temperature for Schwarzschild black hole}\label{fig:GT Sch}
\end{figure}

Since there exists a globally stable phase for the black hole with \Ren entropy, the first-order Hawking-Page (HP) phase transition from no-black hole to stable black hole phases occurs.
It is represented as the blue crossed point in Fig.~\ref{fig:GT Sch}.
In this figure, the red solid and black dashed lines are represented as the black hole with \Ren and GB entropies, respectively.  
It is seen that, at low temperatures, the black hole cannot be formed as the phases lie on the green dotted horizontal line. 
Until the temperature is sufficiently high (known as the HP temperature), the black hole is preferred to form as phases lie on the green dotted slanted line.

In addition, one can realize the effect of non-extensivity by interpreting it as a length scale compared to the size of the system, i.e., the black hole's horizon radius. 
The non-extensivity translates to a quantity in the dimension of length as follows:
\ba
    L_\la=\sqrt{\frac{G}{\pi\la}}.\label{L lamb}
\ea
Note also that such a length scale is compatible only when the parameter $\la$ is positive.
By equipping all constants, the non-extensive length scale is expressed as $L_\la=\sqrt{\hbar G/\lt(\pi k_\text{B}c^3\la\rt)}\sim 10^{-35}/\sqrt{k_\text{B}\la}$ m.
This means that $L_\la$ can be small in the order of $10^{-35}$ m as $k_\text{B}\la$ reaches its upper bound, i.e., $k_\text{B}\la\to1$.
If the magnitude of the non-extensive length scale is comparable to that of the Planck length, it may provide a hint of the connection between the non-extensivity in thermodynamics and the quantum nature of spacetime.
It is crucial to emphasize that these length scales are, however, not physically relevant.
They just coincidentally are in the same order of magnitude.
However, it will be investigated (in the following sections) that the non-extensivity must be sufficiently weak in order to achieve the standard evolution of the Universe.
In other words, the strong non-extensivity scenario, i.e., very small $L_\la$, is not preferable in the context of cosmology.

On the other hand, as $k_\text{B}\la$ goes zero, or equivalently, approaching the GB limit, the non-extensive length scale is infinite.
Based on GB statistics, this is an alternative viewpoint of the Schwarzschild black hole's instabilities.
According to the stability conditions~\eqref{loc stab cond} and \eqref{glob stab cond}, it is obviously seen that the globally stable black hole requires a larger radius than the local one.
In other words, the black hole radius-to-non-extensive length ratio, i.e., $r_h/L_\la$, can identify the stability of the black hole.
One can conclude that the Schwarzschild black hole described by \Ren statistics can be locally and globally stable when
\ba
    \frac{r_h}{L_\la}\gtrapprox 1.98.
\ea
If the non-extensive length is not too large, the black hole is always stable.

It is very important to emphasize that, in the aspect of thermodynamic stability, the effect of non-extensivity analyzed via the \Ren entropy on a black hole in asymptotically flat spacetime is very close to the effect of the existence of the cosmological constant analyzed via the GB entropy on a black hole in asymptotically anti-de Sitter (AdS) spacetime \cite{Hawking:1982dh, Kubiznak:2012wp}.
In other words, the stable phase can emerge due to non-extensivity or negative cosmological constant (see Refs.~\cite{Biro:2013cra, Czinner:2015eyk, Czinner:2017tjq} for further remarks).
According to Ref.~\cite{Nakarachinda:2021jxd}, the aforementioned similarity motivates the authors to investigate a cosmological solution via HDE defined from the black hole with \Ren entropy as will be discussed in the next sections.


\section{Holographic Dark Energy from \Ren black hole}\label{sec: hde}

As previously mentioned, the thermodynamic property of a black hole can play a significant role in defining HDE.
In this section, we consider the Schwarzschild black hole with \Ren entropy.
The behaviors of HDE defined from such a black hole will be discussed in the context of dark energy in the Universe.

We assume that the geometry of spacetime is described by Einstein's general relativity, and then there can exist the black hole.
The holographic dark energy is defined as matter with saturated energy corresponding to the energy of the black hole.
Note that such a dark energy is introduced in the ordinary theory of gravitation.
Hence, our approach does not relate to modified gravity theory or modified Friedmann equation as found in the literature \cite{Manoharan:2022qll}.
According to the thermodynamic stabilities under the isobaric process, we are interested in defining the dark energy from the black hole evolved under such a process. 
The change in the black hole's energy can be written in terms of the energy density and volume as
\ba
	\dd E_\text{saturate}&\sim&\rho_\text{HDE}\dd V.
\ea
In the mentioned scenario, the change of the black hole's energy equals that of the thermodynamic enthalpy, or equivalently, mass.
The energy density of the \Ren holographic dark energy (RHDE) is, therefore, given by
\ba
	\rho_\rde
	&\sim&\lt(\frac{\dd M}{\dd V_\R}\rt)_{P_\R}
	\,=\,\lt(\frac{\pa_L M}{\pa_L V_\R}\rt)_{\la}.
\ea
Here, the volume $V$ is interpreted as the thermodynamic volume of the black hole in \Ren description, i.e., $V_\R$ in Eq.~\eqref{V P Ren}.
The horizon radius $r_h$ from the black hole side is replaced by the characteristic length scale or the IR length scale $L$. 
It is very important to note that the form of the energy density of the original model is modified due to the realization of the thermodynamic ``R\'enyi'' volume, not the \Ren entropy.
This is one of the crucial notions in our approach for which the resulting energy density is different from those in Refs.~\cite{Golanbari:2020coz, Manoharan:2022qll}.
It is because the change of the black hole's mass (under the isobaric process) is identical to that described by the BH entropy,
\ba
	\dd M\big|_{P_\R}
	&=&T_\R\dd S_\R
	\,=\,T_\text{H}\dd S_\text{BH}.
\ea
As a result, the energy density of RHDE can be expressed as
\ba
	\rho_\rde&=&3b^2\lt[\frac{3}{16\pi GL_\la^2\lt\{1+\lt(1-\frac{L_\la^2}{L^2}\rt)\ln\lt(1+\frac{L^2}{L_\la^2}\rt)\rt\}}\rt],\label{rho RHDE in L}
\ea
where the non-extensive length scale $L_\la$ was defined in Eq.~\eqref{L lamb}.
The constant $3b^2$ is introduced to recover the original model in the GB limit~\cite{Hsu:2004ri, Li:2004rb}, i.e., $\displaystyle{\lim_{L_\la\to\infty}\rho_\rde=\frac{3b^2}{8\pi G L^2}}$.

Let us analyze the behaviors of the energy density of RHDE in Eq.~\eqref{rho RHDE in L}.
Under the assumption of fixing the pressure, the non-extensive parameter is appropriately treated as a time-independent variable throughout the evolution of the Universe.
Together with the fact that the parameter $b^2$ is just a constant, the dynamical variable of RHDE model is, thus, only the IR length scale $L$.
It is worthwhile to analyze the asymptotic behaviors of the energy density $\rho_\rde$ in different regimes of $L$.
For a very small $L$ or, equivalently, $L\ll L_\la$, one can approximate
\ba
	\rho_\rde\Big|_{L\ll L_\la}
	&\approx&\frac{3b^2}{8\pi G L^2}+\frac{5b^2}{24\pi GL_\la^2}+\frac{13b^2L^2}{432\pi GL_\la^4}+\hdots\label{rho de small L}
\ea
It is seen that the leading-order term is exactly the same as that in the original model. 
Hence, the Universe filled with RHDE in this limit will mainly evolve in a similar way to that of the original model.
Interestingly, the next-order term is a constant corresponding to the existence of the non-extensive effect. 
Note also that such a constant term appears in the energy density of HDE defined from the Schwarzschild-anti de Sitter black hole \cite{Nakarachinda:2022mlz}.
On the other hand, the energy density $\rho_\rde$ in a very large $L$ regime is approximated as 
\ba
	\rho_\rde\Big|_{L\gg L_\la}
	&\approx&\frac{9b^2}{16\pi G L_\la^2(1+x)}-\frac{9b^2(1-x)}{16\pi G L^2(1+x)^2}+\frac{9b^2L_\la^2(5-x+2x^2)}{32\pi G L^4(1+x)^3}+\hdots\label{rho de large L}
\ea
where $x=\ln\lt(L^2/L_\la^2\rt)$.
Even though $x$ is dynamical, the rate of change is suppressed due to the logarithmic function.
According to this approximation, $x$ can be roughly thought of as a positive constant in the order of unity.
As a result, the first term is made constant and dominant over the rest, hereafter the ``approximated constant''.
Consequently, the Universe will be driven to expand with acceleration. 
If the aforementioned scenario occurs at the late time of the cosmic evolution, the Universe will behave similarly to that of the $\Lambda$CDM model or the AdS-HDE model at late time \cite{Nakarachinda:2022mlz}.


\section{Cosmological solution}\label{sec: cosmo}

In this section, the evolution of the Universe with our proposed dark energy model is investigated.  
The domination of the RHDE at the late time is then illustrated with suitable values of the parameters in the model. 
Interestingly, this RHDE can drive the Universe to expand at an agreed rate with observation.

\subsection{Dark Energy Domination}\label{subsec: DE domination}

To study the evolution of the homogenous and isotropic Universe, it is supposed that the dynamics of the Universe are governed by Einstein's general theory of relativity. 
The field equation is given by
\begin{eqnarray}
	G^\mu_{\hspace{.15cm}\nu}=8\pi GT^\mu_{\hspace{.2cm}\nu},\label{Ein eq}
\end{eqnarray}
where $G^\mu_{\hspace{.15cm}\nu}$ and $T^\mu_{\hspace{.2cm}\nu}$ are the Einstein tensor describing the curvature of the spacetime and the energy-momentum tensor existing in the spacetime, respectively. 
For the curvature sector, we have considered the flat Friedmann-Lema\^itre-Robertson-Walker (FLRW) metric,
\begin{eqnarray}
	\text{d}s^2=-\text{d}t^2+a(t)^2\left(\text{d}r^2+r^2\text{d}\theta^2+r^2\sin^2\theta\text{d}\phi^2\right),
\end{eqnarray}
where $a(t)$ is a scale factor characterizing how the Universe evolves. 
For the matter sector, the energy-momentum tensor is chosen to be taken in the perfect fluid form as follows:
\begin{eqnarray}
	T^\mu_{\hspace{.2cm}\nu}=\text{diag}\Big(-\rho(t), p(t), p(t), p(t)\Big).
\end{eqnarray}
Here, $\rho(t)$ and $p(t)$ are the energy density and pressure of the fluid, respectively. 
Moreover, these energy density and pressure are assumed to be linearly related as
\begin{eqnarray}
	p=w\rho,
\end{eqnarray} 
where $w$ is called the equation of state parameter. 
For instance, $w=1/3$ and $w=0$ for the radiation and non-relativistic matter, respectively. 
As a result, the $(0,0)$ and $(i,j)$ components of the Einstein equation~\eqref{Ein eq} are respectively expressed as
\begin{eqnarray}
	3H^2
	&=&8\pi G\lt(\rho_\text{r}+\rho_\text{m}+\rho_\text{de}\rt),\label{Ein 00}\\
	2\dot{H}+3H^2
	&=&8\pi G\lt(-w_\text{r}\rho_\text{r}-w_\text{m}\rho_\text{m}-w_\text{de}\rho_\text{de}\rt),\label{Ein ij}
\end{eqnarray}
where $H=\dot{a}/a$ is the Hubble parameter. 
The dot denotes the derivative with respect to the cosmic time $t$. 
The quantities with subscripts ``r", ``m" and ``de" denote those for the radiation, matter, and dark energy, respectively.
By defining the density parameter as follows:
\ba
	\Omega_i&=&\frac{8\pi G}{3H^2}\rho_i,\label{Omega i}
\ea
Eq. \eqref{Ein 00} is then rewritten as a constraint equation,
\ba
	1&=&\Omega_\text{r}+\Omega_\text{m}+\Omega_\text{de}.\label{constraint}
\ea
In our consideration, we are simply interested in the case of no interaction among species in the Universe. 
Therefore, each species conserves independently and its energy-momentum tensor obeys the conservation equation, $\nabla_\mu T^\mu_{\hspace{.25cm}\nu}=0$ which can be further expressed as
\ba
	\dot{\rho}_i+3H\big(\rho_i+p_i\big)
	&=&\dot{\rho}_i+3H\big(1+w_i\big)\rho_i
	\,=\,0,\label{con eq}
\ea
where $w_i=p_i/\rho_i$ is the equation of state parameter. 
Note that, for a species with the constant equation of state parameter, the energy density is straightforwardly obtained as a monomial function of the scale factor, $\rho\propto a^{-3(1+w)}$. 
By substituting this to Eq.~\eqref{Omega i}, Eq.~\eqref{constraint} is eventually written as
\ba
	1&=&\Omega_\text{r,0}l_H^2e^{-3(1+w_\text{r})\ln a}+\Omega_\text{m,0}l_H^2e^{-3(1+w_\text{m})\ln a}+\Omega_\text{de}.\label{constraint2}
\ea
The quantities with subscript ``0" stand for those at present.
In the above expression, the present-time scale is set to be unity $a_0=1$.
We have introduced the dimensionless Hubble radius as $l_H=H_0/H$.
It is also noticed that the cosmic time $t$ is now scaled as $\ln a$.


In our cosmological model, the characteristic length scale of RHDE $L$ is chosen as the Hubble radius, i.e., $L=H^{-1}$.
It is very important to note that this choice of the length scale does not contain the problem of causality.
From Eq.~\eqref{rho RHDE in L}, the density parameter of RHDE can be defined as
\ba
	\Omega_\rde&=&\frac{8\pi G}{3H^2}\rho_\rde
	\,=\,\frac{3b^2}{2\frac{l_\la^2}{l_H^2}\lt[1+\lt(1-\frac{l_\la^2}{l_H^2}\rt)\ln\lt(1+\frac{l_H^2}{l_\la^2}\rt)\rt]}.\label{Ode}
\ea
For convenience in analyzing, we define the dimensionless non-extensive variable $l_\la=L_\la H_0$.
Let us emphasize that the RHDE model obviously contains two non-dynamical parameters ($b^2$ and $l_\la$) and one dynamical parameter ($l_H$).

The present-time ratio of RHDE is straightforwardly obtained by replacing $l_H=1$, i.e., $\Omega_\rde\big|_{l_H=1}=\Omega_\text{RHDE,0}$.
The quantity $\Omega_\text{RHDE,0}$ is tightly constrained by recent observational data, e.g., Planck 2018 \cite{Planck:2018vyg}.
This implies that one of the non-dynamical parameters can be eliminated and then be written in terms of the known observable $\Omega_\text{RHDE,0}$ and another non-dynamical parameter.
Conveniently, one chooses to eliminate the parameter $b^2$ with the expression 
\ba
    b^2&=&\frac{2}{3}\Omega_\text{RHDE,0}l_\la^2\lt[1+\lt(1-l_\la^2\rt)\ln\lt(1+\frac{1}{l_\la^2}\rt)\rt].\label{b2 in la}
\ea
As a result, in the recent model, there is only one extra free parameter (i.e., $l_\la$) from those in the $\Lambda$CDM model.
In addition, the density parameter of RHDE can recover to that of the $\La$CDM model in a small $l_\la$ limit, i.e., $\displaystyle{\lim_{l_\la\to0}\Omega_\rde=\Omega_\text{RHDE,0}=\text{constant}}$. 
The larger the non-extensive length $l_\la$, the greater the deviation of RHDE from the $\La$CDM model.
It is important to note that the aforementioned limit cannot be reached 
since there exists the bounded from the maximization of the entropic function $k_\text{B}\la<1$ as mentioned in the previous section. 
This leads to the fact that the magnitude of $l_\la$ is constrained to be small as $l_\la>\sqrt{H_0^2\hbar G/(\pi c^5)}\sim10^{-62}$.
Therefore, RHDE can mathematically describe the identical cosmic evolution of the Universe as the standard $\La$CDM model does when $l_\la$ approaches zero.
However, this situation is forbidden due to consistency in \Ren statistics.

\begin{figure}[!ht]
	\includegraphics[scale=0.42]{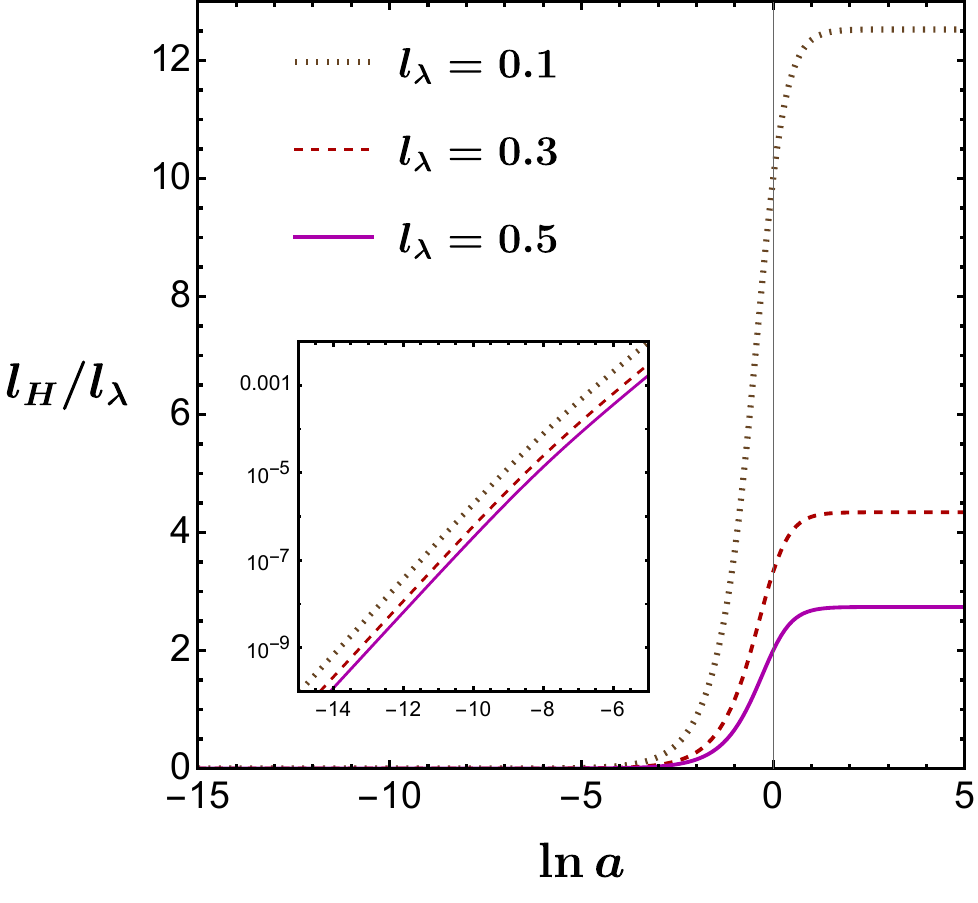}
 	\hspace{1cm}
	\includegraphics[scale=0.47]{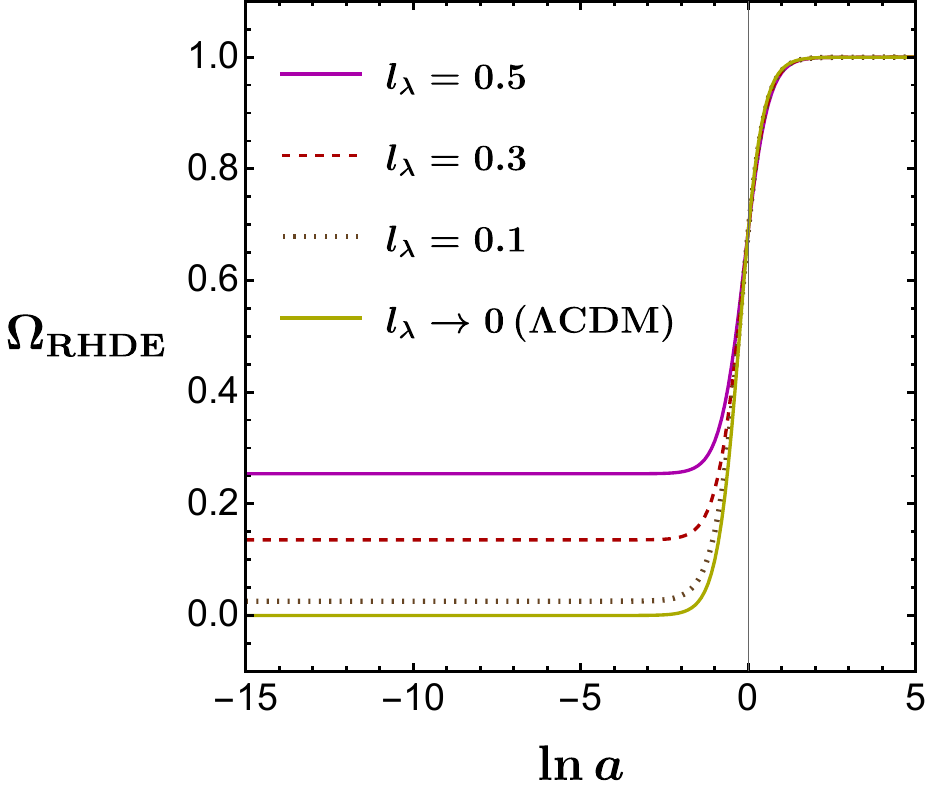}
\caption{Evolutions of ratio $l_H/l_\la$ (left) and RHDE density parameter (right) with respect to $\ln a$ for various values of $l_\la$. In these plots, we set  $\Omega_\text{RHDE,0}=0.69$.}\label{fig:lH Ode}
\end{figure}

For a given non-extensive length scale $l_\la$, the evolution of the parameter $l_H$ is evaluated by solving the constraint equation~\eqref{constraint2} together with Eqs.~\eqref{Ode}~and~\eqref{b2 in la}.
The numerical result is illustrated in the left panel of Fig.~\ref{fig:lH Ode}.
According to the evolution, the magnitude of $l_H$ (or $H^{-1}$) compared to $l_\la$ (or $L_\la$) is very small in the range $\ln a<-5$.
In this regime, the approximation in Eq.~\eqref{rho de small L} appropriately explains the behavior of RHDE.
Hence, RHDE acts as the scaling solution in the early time.
The standard cosmic evolution corresponding to those of the radiation- and matter-domination epoch is obtained for $\ln a<-5$.
As time evolves, the left panel of Fig.~\ref{fig:lH Ode} also shows that the magnitude of $l_H$ gets larger and larger until it equals the magnitude of $l_\la$ at $-2\lessapprox\ln a\lessapprox0.5$.
Note that the time that $l_H=l_\la$ depends on the non-extensivity.
For a sufficiently small $l_\la$, the magnitude of $l_H$ can be much greater than that of $l_\la$ in a regime $\ln a>0$.
The approximation in Eq.~\eqref{rho de large L} is applicable to describe RHDE at the late time.
RHDE with a suitable $l_\la$ can provide the dark energy-dominated epoch due to its approximated constant term in energy density, i.e., the first term in the right-hand side of Eq.~\eqref{rho de large L}, being dominant.
The results are confirmed in the right panel of Fig.~\ref{fig:lH Ode}.
RHDE will be dominant over other species at the late time.

\begin{figure}[!ht]
	\includegraphics[scale=0.32]{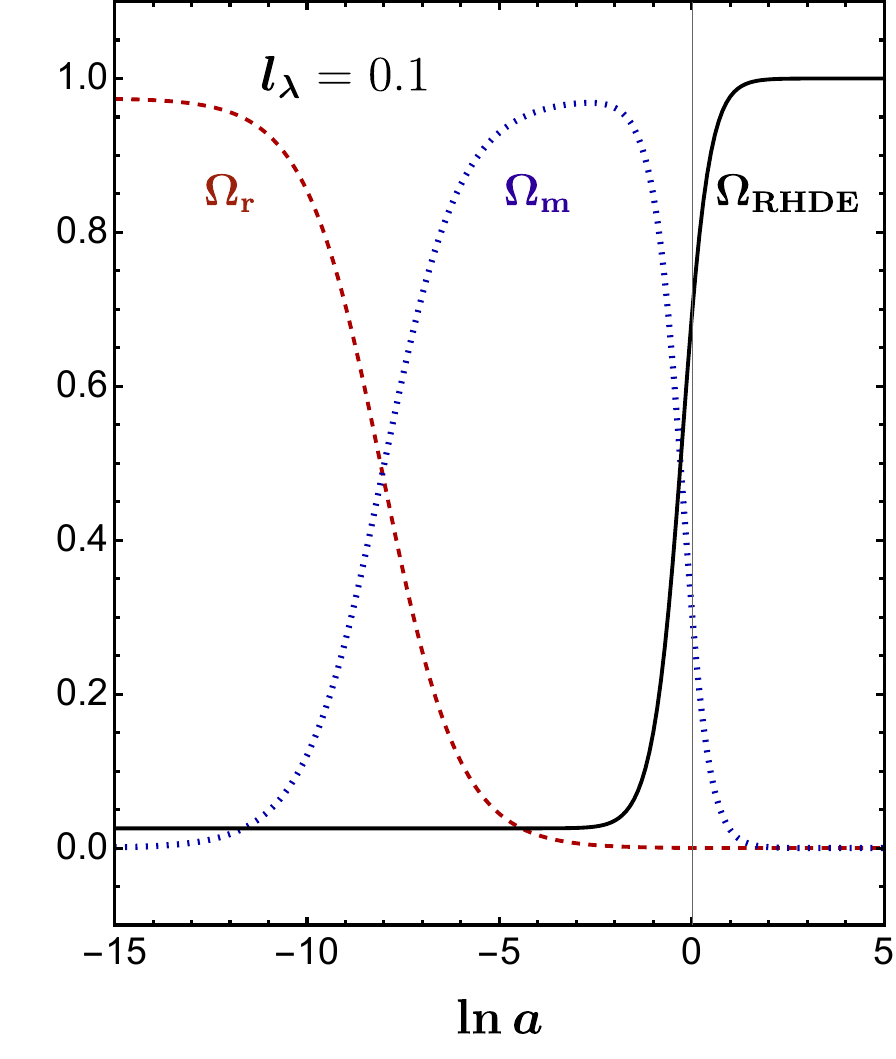}
	\hspace{0.5cm}
	\includegraphics[scale=0.32]{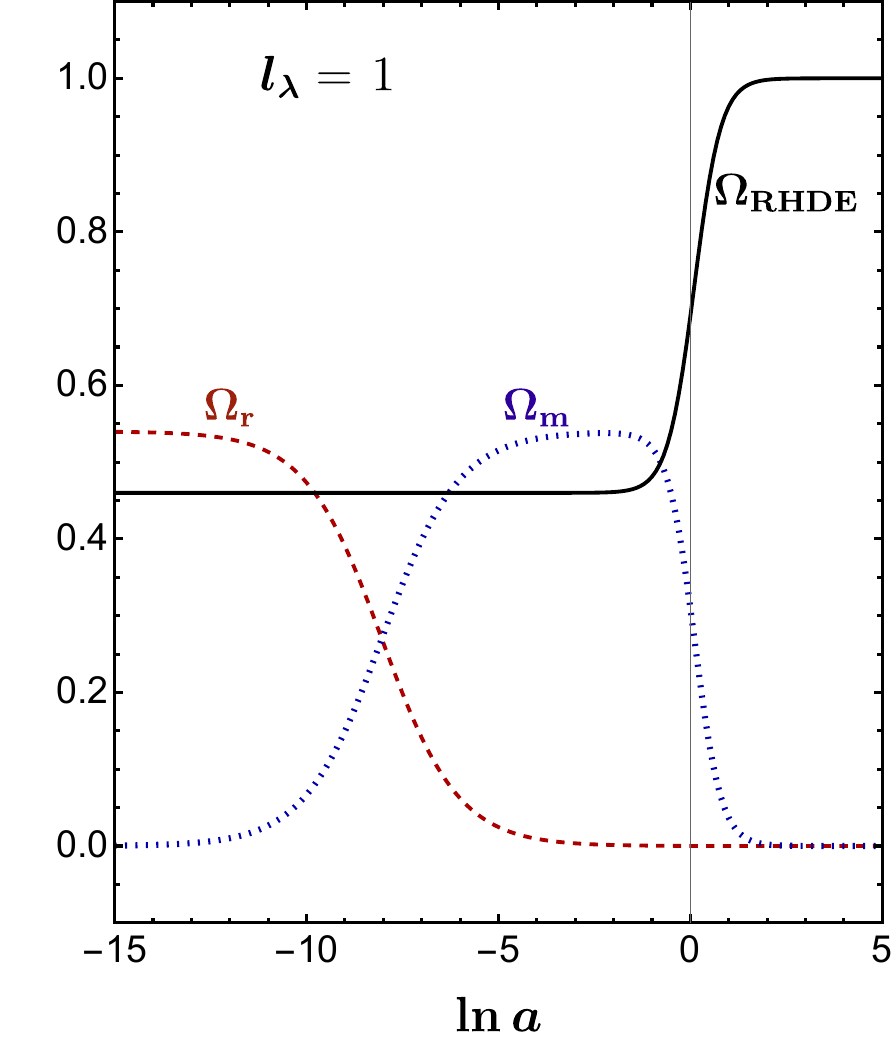}
	\hspace{0.5cm}
	\includegraphics[scale=0.32]{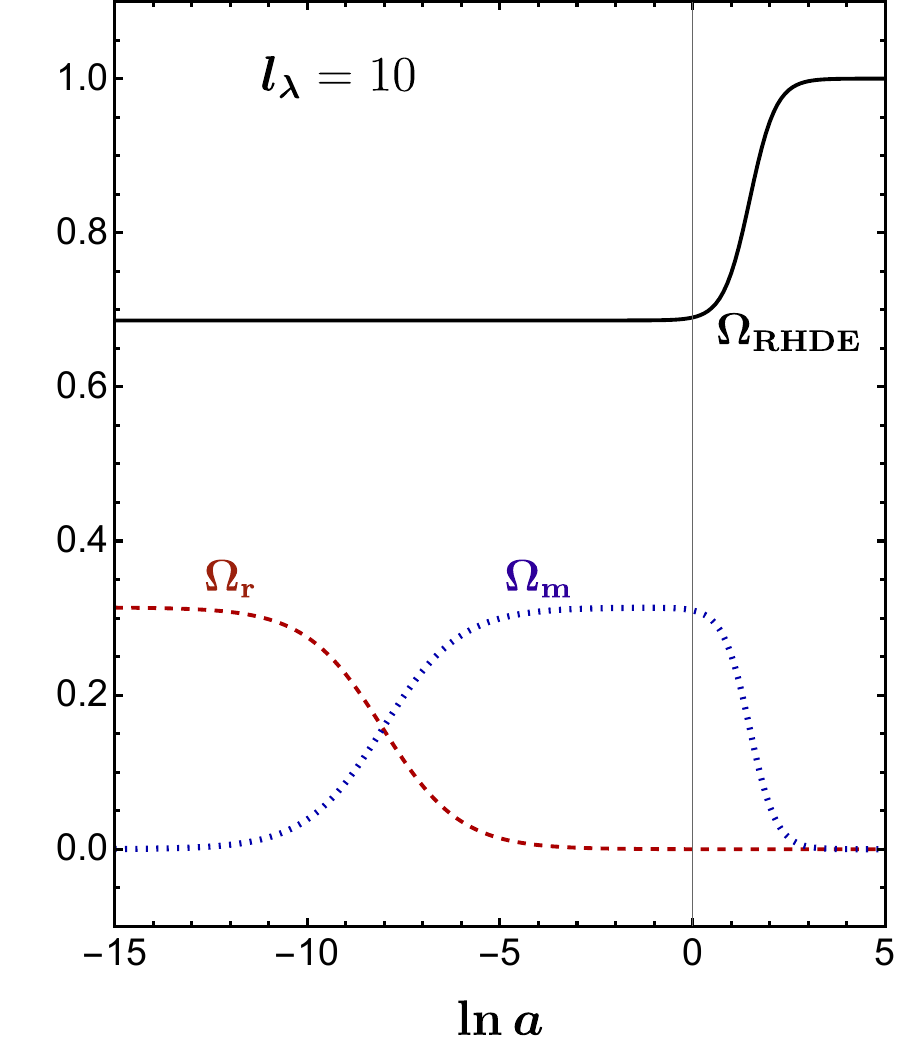}
\caption{Evolution of density parameters for all species with respect to $\ln a$ for $l_\la=0.1$ (left), $l_\la=1$ (middle) and $l_\la=10$ (right). Note that we have set the value of the present-time density parameters for radiation and matter as $\Omega_\text{r,0}=10^{-4}$ and $\Omega_\text{m,0}=1-\Omega_\text{r,0}-\Omega_\text{RHDE,0}$, respectively.}\label{fig:all O evo}
\end{figure}

Another feature of RHDE is that the non-extensive length scale controls the ratio of the dark energy in the early times as shown in the right panel of Fig.~\ref{fig:lH Ode} or Fig.~\ref{fig:all O evo}.
It can be realized by considering the density parameter in a limit of small $l_H$ as follows:
\ba
	\lim_{l_H\to0}\Omega_\rde
	=\frac{2}{3}l_\la^2\Omega_\text{RHDE,0}\lt[1+(1-l_\la^2)\ln\lt(1+\frac{1}{l_\la^2}\rt)\rt].
\ea 
If the non-extensive length scale is too large, the ratio of radiation will be too low.
Thus, there is not enough production of fundamental particles' nuclei.
Such a strong constraint is known as the Big Bang Nucleosynthesis (BBN) constraint which should be satisfied for the early dark energy.
This constraint provides that $\Omega_\text{de}<0.045$ at the radiation-dominated epoch \cite{Bean:2001wt}.
It leads to a maximum bound in $l_\la$ as $l_\la<0.142$ when $\Omega_\text{RHDE,0}=0.69$.
In fitting the model with low-redshift observational data sets (in Sec.~\ref{sec: obs}), the range of $l_\la$ would appear to be limited as $0<l_\la<0.2$.
It turns out that the non-extensive length scale $L_\la$ is in the order of $10^{25}$ m which is equivalent to $k_\text{B}\la\sim10^{-60}$.
It is determined how much non-extensiveness should be required in order to deviate from standard GB thermodynamics.
As mentioned in Sec.~\ref{sec: bh}, this is the sufficient weakness of non-extensivity investigated from our cosmological model.

As discussed in this subsection, we have shown that the RHDE can be dominant over other species in the Universe. The next aim of this work is to investigate how the Universe fulfilled with the RHDE evolves. In other words, we are interested in determining whether the proposed dark energy can drive the accelerated expansion of the Universe at a suitable rate.


\subsection{Accelerated Expansion}\label{subsec: acc expansion}

\begin{figure}[!ht]
	\includegraphics[scale=0.4]{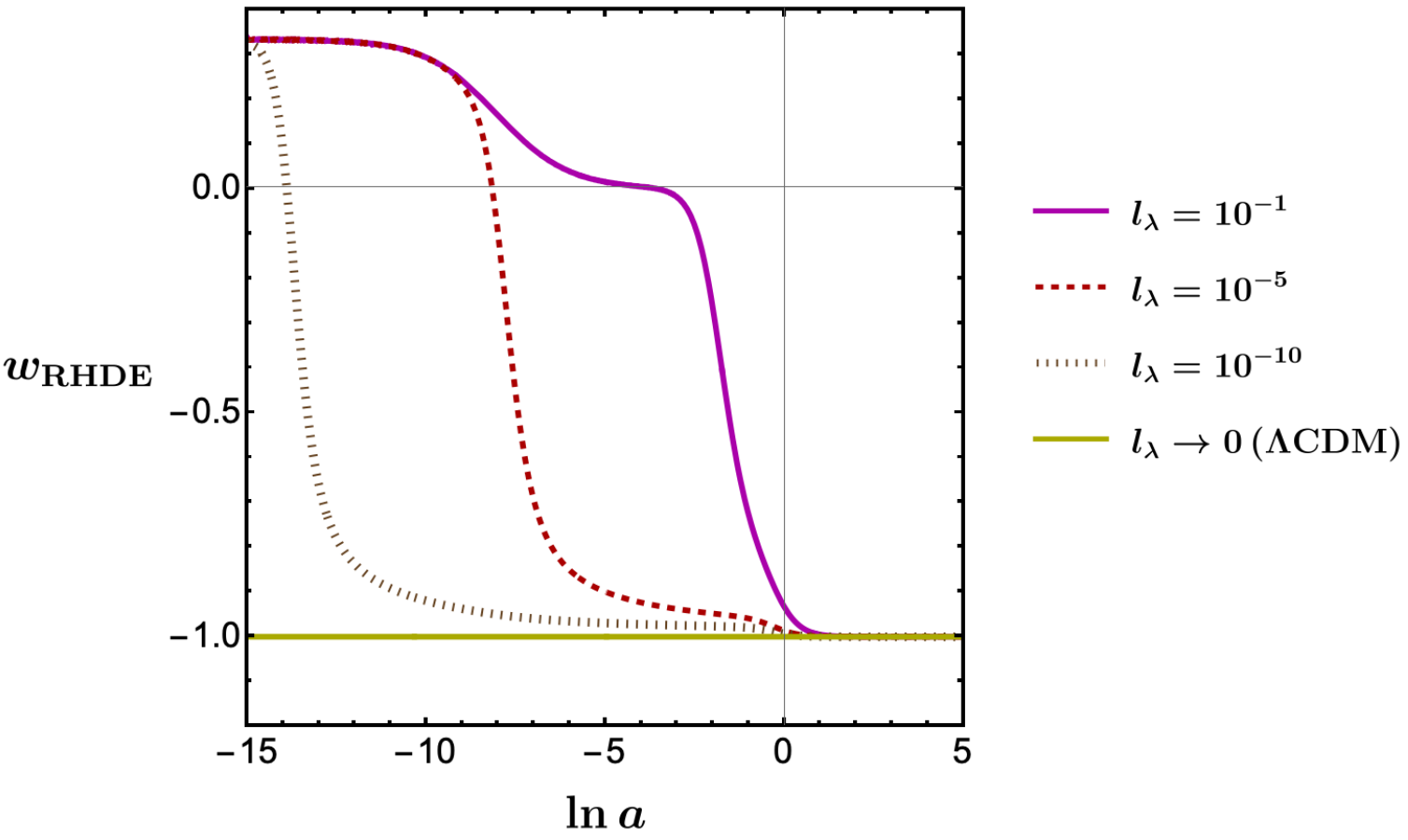}
\caption{Equation of state parameter of RHDE with respect to $\ln a$ for various values of $l_\la$}\label{fig:wde}
\end{figure}

\begin{figure}[!ht]
	\includegraphics[scale=0.4]{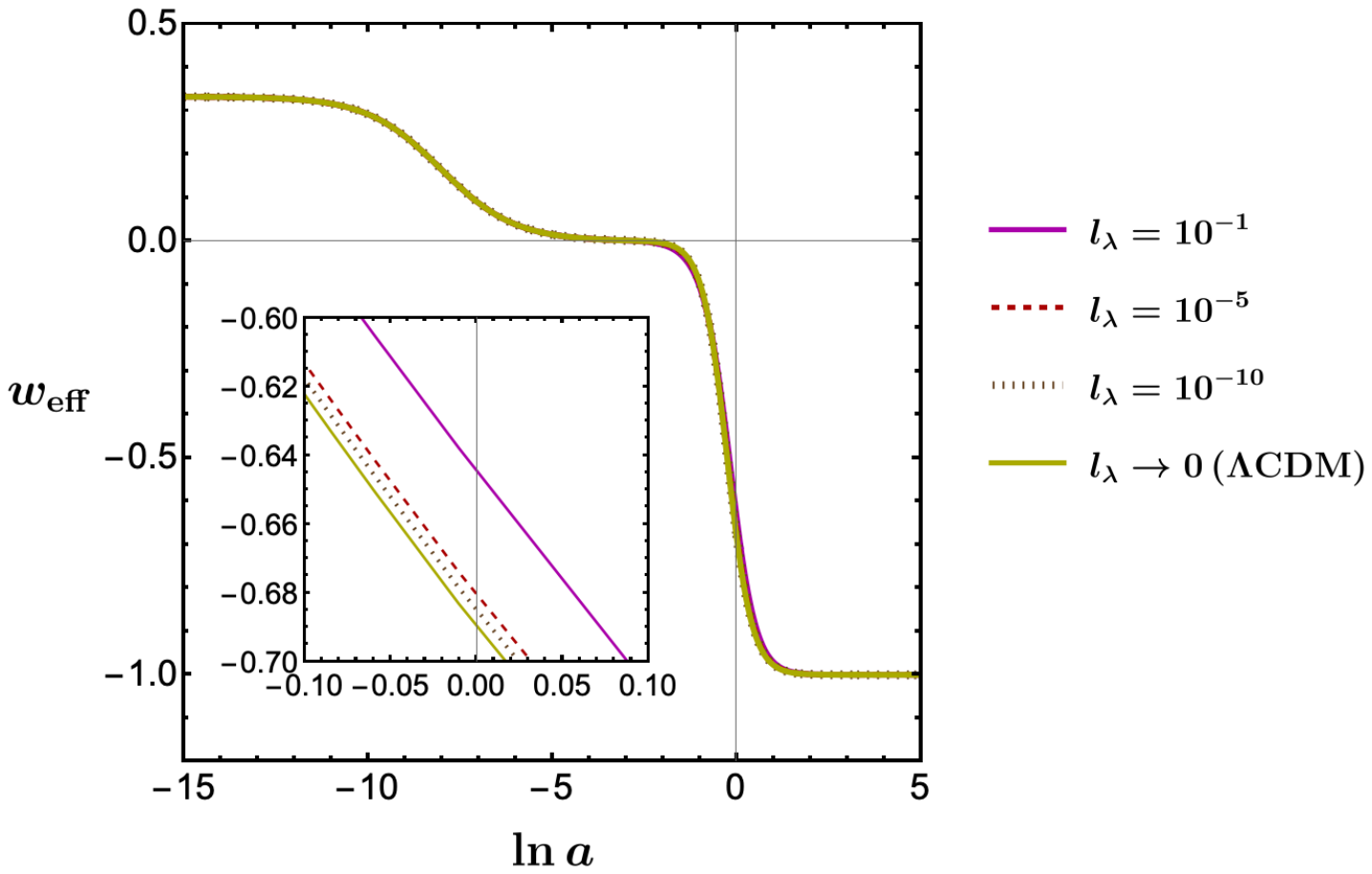}
\caption{Effective equation of state parameter with respect to $\ln a$ for various values of $l_\la$}\label{fig:weff}
\end{figure}

According to the conservation equation \eqref{con eq}, the equation of state parameter for RHDE is given by
\ba
	w_\rde
	&=&-1-\frac{\dot{\rho}_\rde}{3H\rho_\rde}
    \,=\,-1+\frac{\frac{\dd}{\text{d}(l_H^2)}\rho_\rde}{\rho_\rde}l_H^4\lt(\frac{2\dot{H}}{3H_0^2}\rt).
\label{w de}
\ea
This quantity is essential for the dark energy model because it characterizes the rate of the expansion of the Universe. For the effective evolution of the Universe for all epochs, one can determine the effective equation of state parameter using Eqs.~\eqref{Ein 00} and \eqref{Ein ij},
\ba
	w_\text{eff}&=&\frac{\sum_ip_i}{\sum_i\rho_i}
	\,=\,-1-\frac{2\dot{H}}{3H^2}
	\,=\,-1-l_H^2\lt(\frac{2\dot{H}}{3H_0^2}\rt).
\label{w eff}
\ea
Obviously, there is still an unknown quantity $2\dot{H}/(3H_0^2)$.
It can be evaluated by subtracting Eq.~\eqref{Ein 00} by \eqref{Ein ij} and using Eqs.~\eqref{w de}.
One finds,
\ba
	\frac{2\dot{H}}{3H_0^2}
	&=&\frac{(1+w_\text{r})\Omega_\text{r}+(1+w_\text{m})\Omega_\text{m}}{l_H^2\lt[1+l_H^2\frac{\frac{\dd}{\text{d}(l_H^2)}\rho_\rde}{\rho_\rde}\lt(1-\Omega_\text{r}-\Omega_\text{m}\rt)\rt]}.\label{H dot}
\ea
From the fact that the energy density $\Omega_\rde$ is a function of $l_H$, both equation of state parameters $w_\rde$ and $w_\text{eff}$ are, therefore, described by the parameter $l_H$ for each $\ln a$. 
The numerical results are shown in Figs.~\ref{fig:wde} and \ref{fig:weff}. 

According to the approximation in Eq.~\eqref{rho de small L}, the energy densities at early time and late are, respectively, given by
\ba
    \rho_\rde\big|_{l_H\ll l_\la}&\propto&l_H^{-2}.
\ea
Applying it to Eq.~\eqref{H dot}, the equation of state parameter in Eq.~\eqref{w de} can be expressed as that of the scaling solution as previously mentioned,
\ba
    w_\rde&\approx&\frac{w_\text{r}\Omega_\text{r}+w_\text{m}\Omega_\text{m}}{\Omega_\text{r}+\Omega_\text{m}}.\label{wde approx early}
\ea
This scaling behavior is valid in the radiation domination epoch and the beginning phase of the matter domination epoch when the non-extensive length scale is sufficiently large.
For example, the dynamic of RHDE with $l_\la=10^{-1}$ represents as the solid magenta line in Fig.~\ref{fig:wde}.
It is also noticed that the period at which RHDE behaves as the scaling solution decreases as $l_\la$ gets smaller.
Instead, RHDE with smaller $l_\la$ gets closer to the cosmological constant because the magnitude of $l_H$ is not sufficiently small.
It means the approximations in Eq.~\eqref{rho de small L} or Eq.~\ref{wde approx early} are not applicable in describing the RHDE at the early time.
Even though the RHDE with very small $l_\la$ can occur, it does not have a significant effect on the standard cosmic evolution (as seen from $w_\text{eff}$ in Fig.~\ref{fig:weff}).
This is because the ratio of RHDE (represented by $\Omega_\rde$) is also very small in such a case.

For another asymptotic limit corresponding to Eq.~\eqref{rho de large L}, the energy density is approximately constant at the late time.
The equation of state parameter of RHDE approaches $w_\rde\to-1$ as seen in Fig.~\ref{fig:wde}.
Therefore, this model of dark energy can drive the Universe at the late time to expand with the correct rate of acceleration.


\section{Observational Constraints}\label{sec: obs}

\begin{figure}[!ht]
\begin{center}
	\includegraphics[scale=0.44]{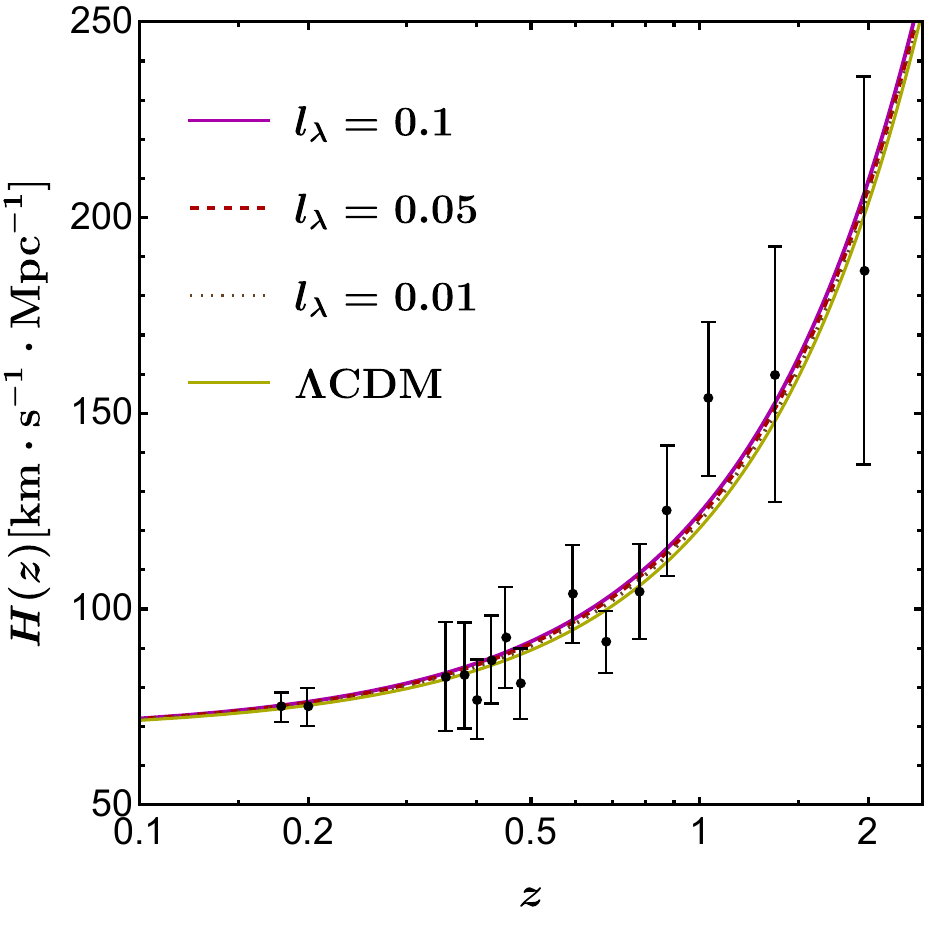}
    \hspace{1.cm}
	\includegraphics[scale=0.295]{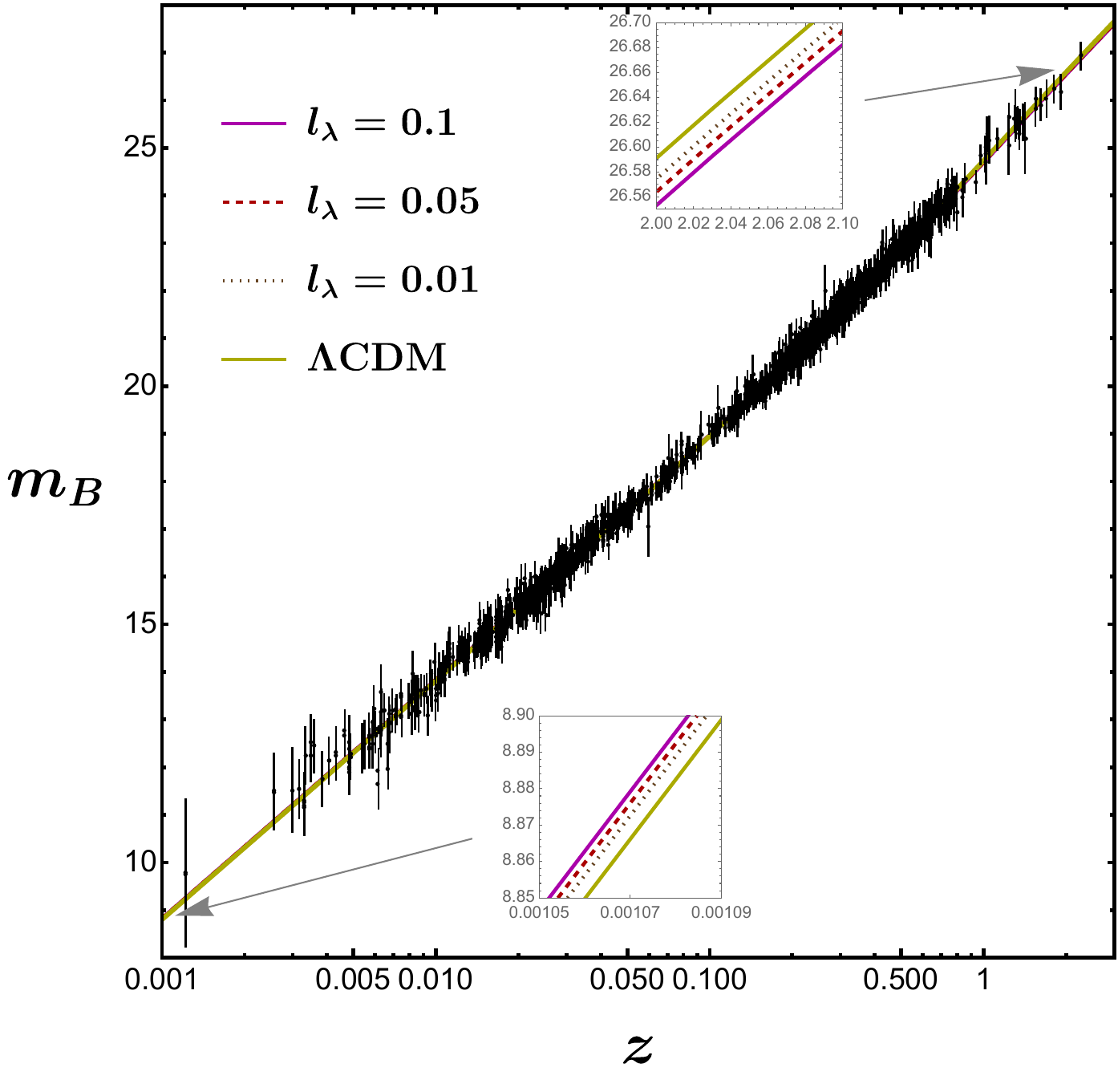}
\end{center}
\caption{Comparison of observables predicted from RHDE model for various values of $l_\la$ to cosmic chronometer data set \cite{Moresco:2020fbm} (left) and Pantheon+ data set \cite{Brout:2022vxf} (right)}
\label{fig: obs plots}
\end{figure}

In this section, we will fit the RHDE model to the low-redshift observational data sets.
As discussed in the previous section, the dynamics of the Universe can be affected due to the existence of non-extensivity. 
For example, the left panel of Fig.~\ref{fig: obs plots} shows the Hubble parameter $H$ for low redshift $z=(1-a)/a$. 
In the right panel, we plot the apparent magnitude $m_B$ which is defined as $\displaystyle{m_B=5\log_{10}\lt(\frac{d_L}{\text{Mpc}}\rt)+25+M}$, where $\displaystyle{d_L=c(1+z)\int_0^z\frac{\dd z'}{H(z')}}$ is the luminosity distance and $M$ is the absolute magnitude introduced as a fitting parameter.
It is seen that the modification in cosmic evolution due to varying $l_\la$ affects the values of the observables predicted from the model at each redshift.
Accordingly, these plots suggest that the model parameter $l_\la$ must be fitted by the observations.

In this work, the likelihood analysis is performed by using the Markov Chain Monte Carlo (MCMC) code \texttt{SimpleMC}.\footnote{It is publicly available in \href{https://igomezv.github.io/SimpleMC/index.html}{https://igomezv.github.io/SimpleMC/index.html}}
To investigate the best-fit model parameter, the considered observational data sets are listed as follows:
\begin{itemize}
    \item Cosmic chronometers (CC) with covariance matrix (2020) \cite{Moresco:2020fbm}
    \item 1550 distinct type Ia supernova (SNIa) containing 1701 light curves from Pantheon+ compilation \cite{Brout:2022vxf}
    \item Recent baryon acoustic oscillation (BAO) data from SDSS-IV DR16 \cite{SDSS:2000hjo, Percival:2007yw, Chuang:2013hya, Dawson:2015wdb, eBOSS:2020yzd}
\end{itemize}
To create samples for MCMC analysis, the Metropolis-Hastings algorithm was used. Additionally, we assess the quality of fits where an increasing number of parameters is penalized by using the Akaike Information Criterion (AIC). With the $k$ parameters and the highest Likelihood $\mathcal{L}_{\rm max}$, AIC is defined as
\ba
    {\rm AIC} = -2\ln \mathcal{L}_{\rm max} + 2k.\label{AIC}
\ea
Notice that the last term is introduced in order to penalize and prevent the over-fitting issue. When comparing two models, one can utilize the disparity in the estimators to support the models' statistical compatibility. By applying Jeffreys' scale \cite{1939thpr.bookJ}, statistical compatibility is indicated by a value of $\Delta$AIC $\leq 2$, and mild tension between models is indicated by a criterion of $2 < \Delta$AIC $\leq 5$. When $\Delta$AIC $> 5$, there is a large statistically significant tension between the models. To verify convergence, the Gelman-Rubin diagnostic (GRstop) has been employed.

\begin{figure}[!ht]
\begin{center}
	\includegraphics[scale=0.5]{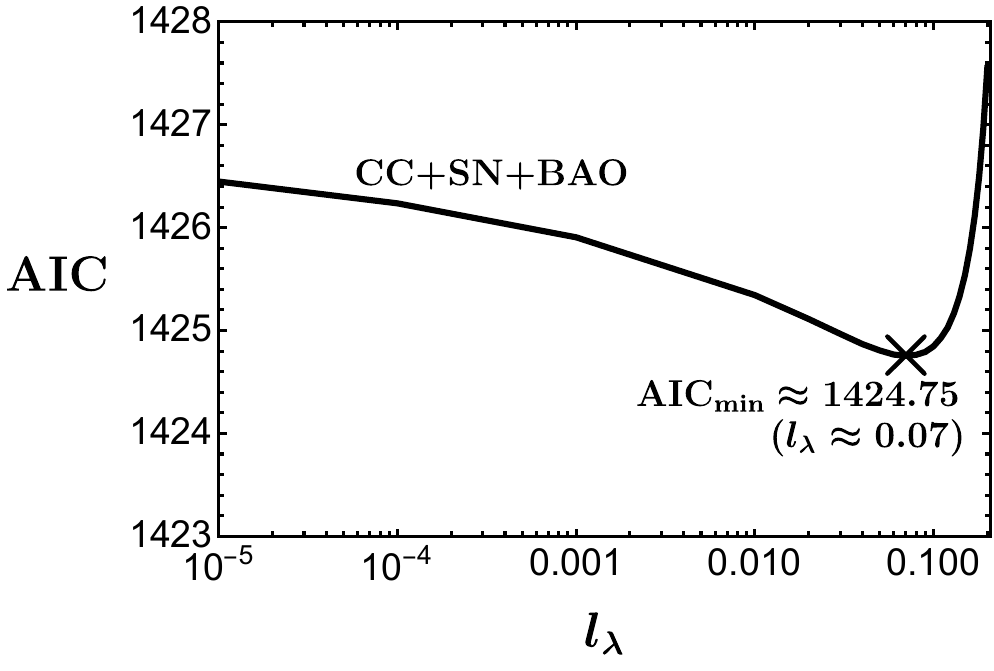}
	\\\vspace{0.5cm}
	\includegraphics[scale=0.4]{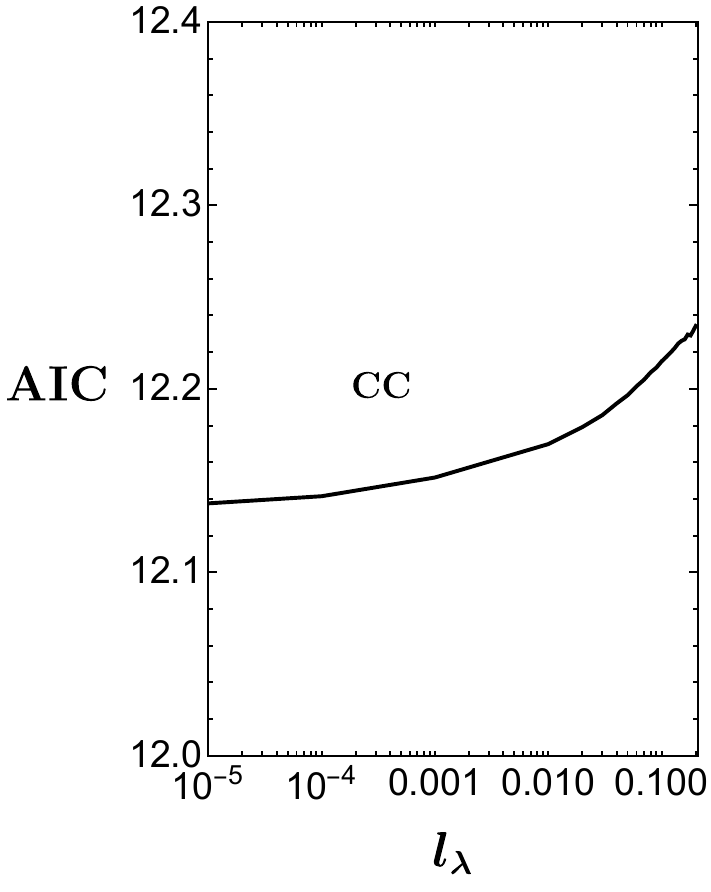}\quad
	\includegraphics[scale=0.4]{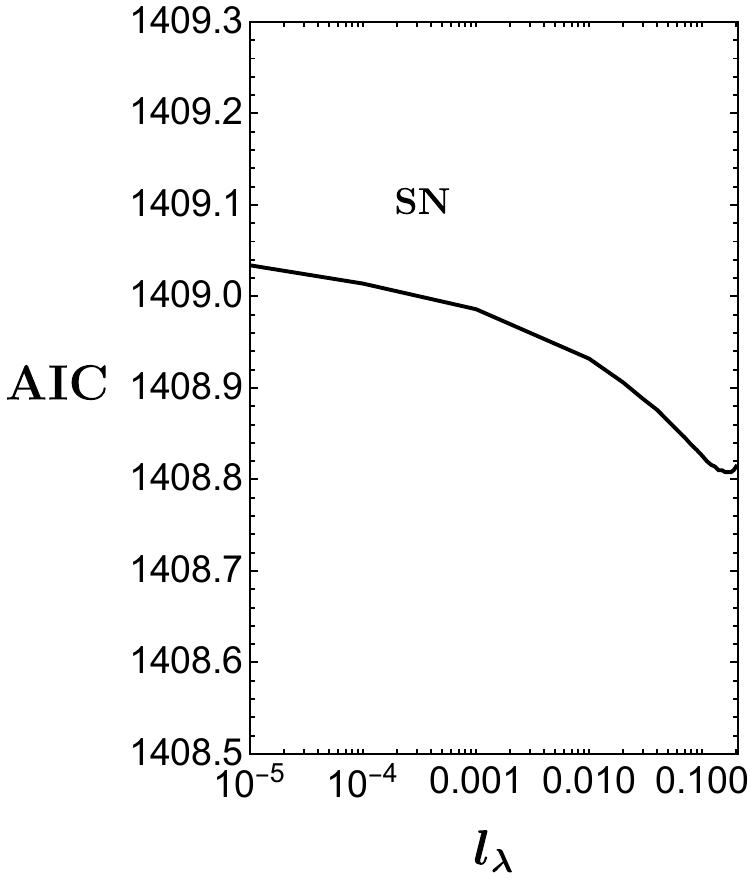}\quad
	\includegraphics[scale=0.4]{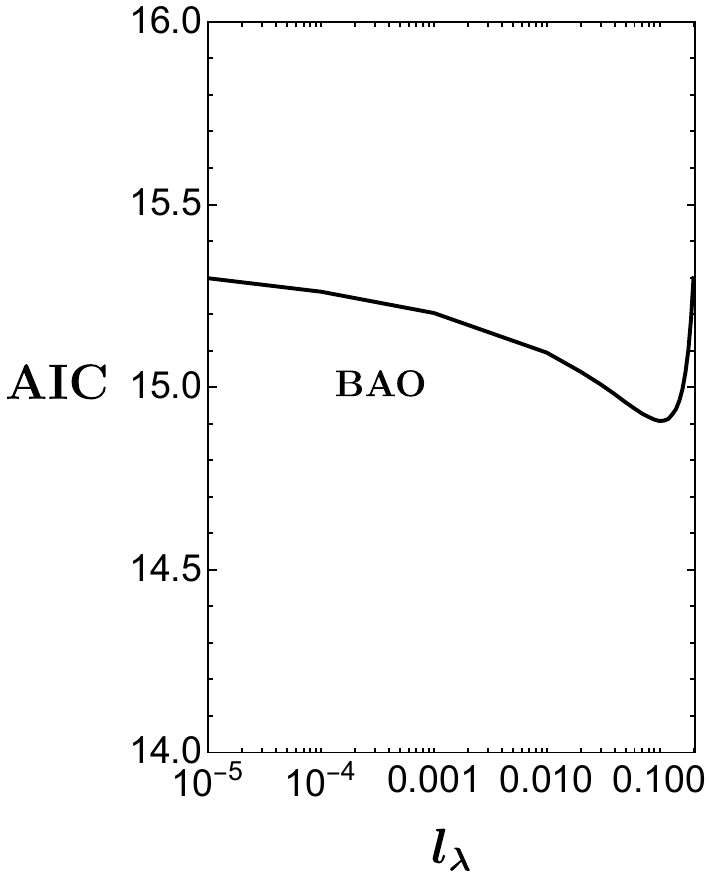}\quad
\end{center}
\vspace{-0.5cm}
\caption{
AIC versus values of $l_\la$ in fitting RHDE model with all observational data sets (top), only CC (left-bottom), Pantheon+ (middle-bottom), and BAO (right-bottom) data sets}
\label{fig: AIC vs ll}
\end{figure}

\begin{figure}[!ht]
\begin{center}
	\includegraphics[scale=0.36]{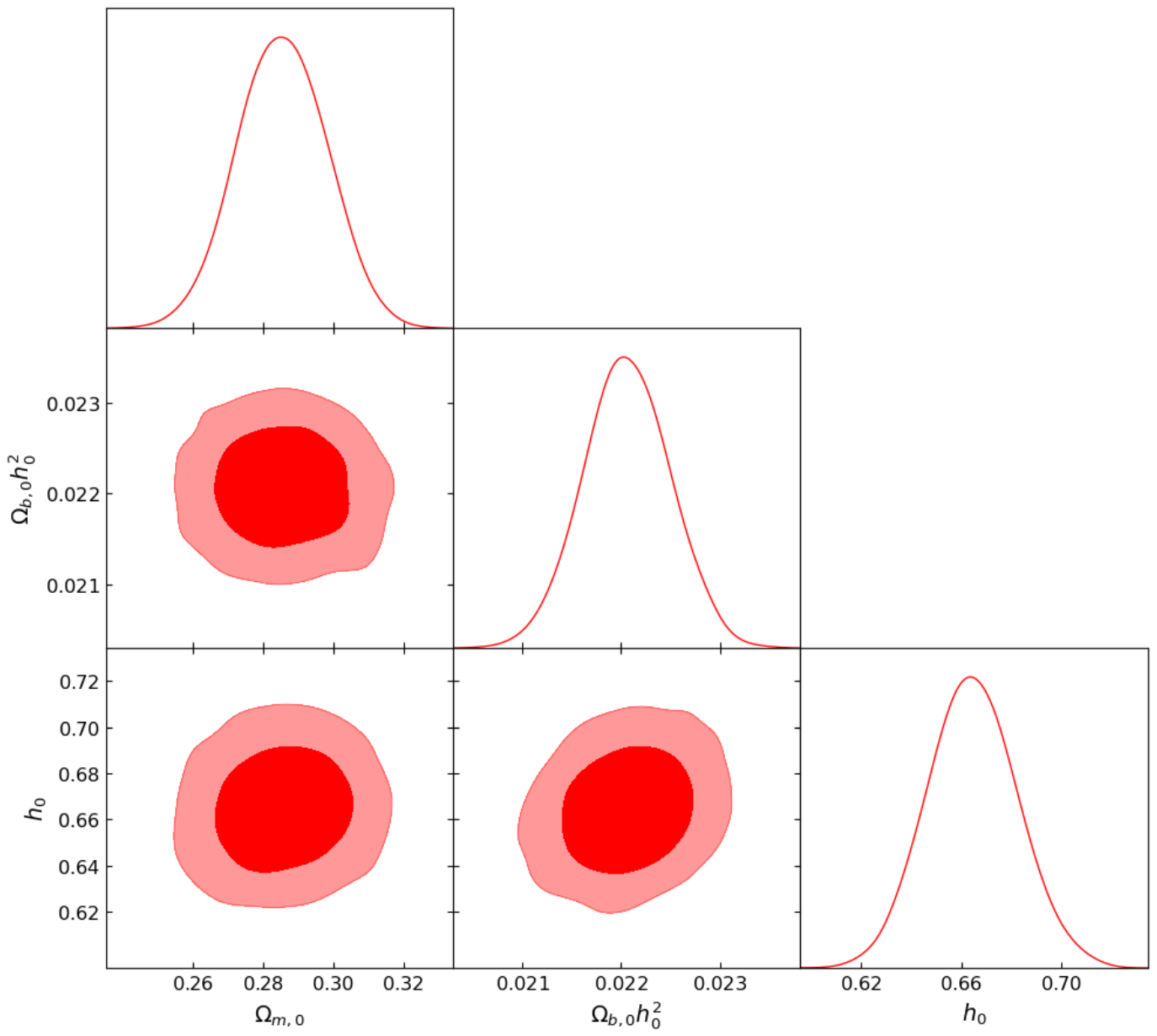}
\end{center}
\caption{Two-dimensional 68\% and 95\% CL allowed regions and the probability posterior distributions for model parameters from RHDE model with  $l_\la\approx0.07015$ (it gives $\text{AIC}_\text{min}$) for the parameter space $\{\Omega_{m,0}, \Omega_{b,0}, h_0\}$}
\label{fig: contour}
\end{figure}

Unfortunately, it is complicated to do the statistical analysis by keeping $l_\la$ as a free parameter. This is because the Hubble parameter cannot be written as an explicit function of $l_\la$.
We instead choose to consider the parameter space: \{$\Omega_\text{m,0}, \Omega_\text{b,0}, h_0=H_0/100$\} for a given $l_\la$, and then evaluate the value of $l_\la$ which gives the lowest AIC. 
The results of AIC with respect to $l_\la$ are illustrated in Fig.~\ref{fig: AIC vs ll}.
It is found that the existence of the non-extensive effect is not preferable by CC data. 
In other words, RHDE with $l_\la\to0$, or equivalently $\La$CDM is the best-fit model for CC data as seen in the left-bottom panel of Fig.~\ref{fig: AIC vs ll}. 
In contrast, from the result shown the middle-bottom panel of Fig.~\ref{fig: AIC vs ll}, SNIa data seems to prefer a large non-extensive length scale $l_\la\approx0.164$.
Such a case indeed contradicts the bound from the BBN constraint.
The two data sets, however, give less contribution in constraining the model parameter.
The most significant effect on AIC in fitting the model comes from BAO data which prefers the intermediate value of $l_\la\approx0.102$ (see the right-bottom panel of Fig.~\ref{fig: AIC vs ll}).
As a result, in fitting RHDE with all data sets, AIC is minimized as $\text{AIC}_\text{min}\approx1424.754$ when $l_\la\approx0.07015$.
The observational analysis for $l_\la$ giving $\text{AIC}_\text{min}$, the probability posterior distribution of parameters and their central values are reported in Fig.~\ref{fig: contour} and the second row of Tab.~\ref{Tab: xxx}, respectively.
\begin{table}[h!]
\centering
\caption{Observational constraints on the parameters of RHDE, AdS-HDE and $\La$CDM using CC, Pantheon+, and BAO data sets}
\begin{tabular}{|l|c|c|c|c|c|c|}
	\hline
Models & $\Omega_\text{m,0}$ & $\Omega_\text{b,0}h^2$ & $h_0$ & $l_\la$ &  $\Omega_{R,0}$ & $b^2$  
    \\\hline\hline
RHDE with $l_{\la,\text{AIC}_\text{min}}$ & $\,0.2856^{+0.0130}_{-0.0125}$\, & \,$0.0221^{+0.0004}_{-0.0004}$\, & \,$0.6643^{+0.0178}_{-0.0174}$\, & \,0.07015\, & -- & 0.0144\,
    \\\hline
AdS-HDE: fix $b^2$ & $\,0.3128^{+0.0127}_{-0.0125}$\, & \,$0.0220^{+0.0004}_{-0.0005}$\, & \,$0.6978^{+0.0195}_{-0.0188}$\, & -- &  \,$0.0036^{+0.0044}_{-0.0029}$\, & 1
    \\\hline 
AdS-HDE: & $\,0.3124^{+0.0136}_{-0.0131}$\, & \,$0.0220^{+0.0005}_{-0.0005}$\, & \,$0.6996^{+0.0194}_{-0.0191}$\, & -- & \,$0.0043^{+0.0050}_{-0.0028}$\, & \,$0.8442^{+0.4069}_{-0.2151}$\,
    \\\hline     
$\Lambda$CDM & $0.3121^{+0.0134}_{-0.0130}$ & $0.0220^{+0.0005}_{-0.0005}$ & $0.6970^{+0.0194}_{-0.0188}$ & -- & --& --
    \\\hline	
\end{tabular}\label{Tab: xxx}
\end{table}

By comparing to other models, the values of AIC for each model are shown in Tab.~\ref{Tab: AIC}. 
\begin{table}[h!]
\centering
\caption{Estimator of each model compared with $\La$CDM}
\begin{center}
\begin{tabular}{|l|c|c|}
	\hline
\hspace{1cm}Models & AIC & $\Delta$AIC \\ 
 	\hline\hline	
RHDE with $l_{\la,\text{AIC}_\text{min}}$ & \,\,1424.754\,\, & \,\,$-2.632$\,\, \\
 	\hline  
AdS-HDE: fixed $b^2$ & 1429.912 & 2.526 \\
 	\hline
AdS-HDE & 1431.910 & 4.524 \\
 	\hline	
$\La$CDM & 1427.386 & 0 \\
 	\hline
\end{tabular}
\end{center}\label{Tab: AIC}
\end{table}
Note that AdS-HDE in Tabs.~\ref{Tab: xxx}~and~\ref{Tab: AIC} represents the model with the characteristic length scale as the particle horizon \cite{Nakarachinda:2022mlz}.
To compare RHDE and AdS-HDE models, even though both can explain the late-time expansion equally well, the RHDE is preferred by observations since it has a lower value of AIC.
To compare with $\Lambda$CDM in Tab.~\ref{Tab: AIC} the RHDE with fixed $l_\la$ performs fits with experimental data statistically better than $\Lambda$CDM with the mild tension.
However, the non-extensive length scale $l_\la$ should be treated as an additional parameter in the parameter space.
Therefore, the value of AIC for RHDE in Tab.~\ref{Tab: AIC} should be added by two due to the increase in the number of parameters in Eq.~\ref{AIC}. With this consideration, the difference of their AIC values falls instead in the range $\Delta\text{AIC}\leq2$ implying that RHDE and $\La$CDM models are statistically compatible.

One of the other crucial analyses for dark energy models is the cosmological perturbation analysis.
Unfortunately, it is difficult to investigate the full perturbation analysis for the holographic dark energy model. 
This is because the field description of this type of dark energy is ambiguous so the general form of the dark energy perturbation cannot be explicitly determined. 
In computing the relevant quantities characterizing instability, such as the effective and adiabatic sound speeds, it is necessarily specified by hand, not a full analysis, and may not be useful to properly analyze observational constraints, e.g., CMB data.
Therefore, the perturbation part is excluded from the likelihood analysis, and anticipated that our analysis is sufficient to constrain the models with observations.


\section{Conclusion}\label{sec: conclu}

In this work, the dark energy was defined from the thermodynamic properties of the Schwarzschild black hole with a non-extensive effect.
Such non-extensivity has been analyzed by treating the entropy as the logarithmic function of the area. 
This consideration is associated with the \Ren description from a thermodynamic point of view.
The non-extensive effect contributed from \Ren entropy can significantly stabilize the black hole similar to the effect of the negative cosmological constant on the anti-de Sitter black hole. 
It also provides an alternative aspect in identifying the stable phase of the black hole by comparing the radius of the black hole to the non-extensive length scale $L_\la$.
The global stability requires larger $L_\la$ than the local one does (see Eqs.~\eqref{loc stab cond} and \eqref{glob stab cond}).
The non-extensive thermodynamics will be reduced to the standard GB one by taking this length scale to infinity.
Hence, the black hole described by GB statistics is always unstable.
As motivated by the stability of the black hole, the energy density of the holographic dark energy can be obtained from the change of enthalpy under the isobaric process.
Remarkably, such dark energy was introduced as the additional species on the spacetime described by general relativity so we did not deal with the modified gravity theory.

It was interestingly found that the energy density of RHDE contains the approximated constant term.
This term can be dominant in the late-time evolution of the Universe.
Therefore, it can drive the expansion with the corrected rate of acceleration in the same way as the cosmological constant in the $\La$CDM model does.
One of the important remarks is that the non-extensive length scale $l_\la$ is only one additional parameter to the standard $\La$CDM model.
By fitting the proposed RHDE model to several observations, we use the likelihood analysis.
The lowest AIC is an estimator in identifying which value of $l_\la$ is preferred.
In the recent procedure, the parameter space is kept the same as that of the $\La$CDM model.
The result showed that the minimum AIC was obtained when the non-extensive length scale is $l_\la\approx0.07$ (or $L_\la\approx0.07H_0^{-1}$).
It can be concluded that the RHDE and $\La$CDM models are not different in the quality of fit of the observational data sets from a statistical point of view.
Moreover, both of them are more favorable than the AdS-HDE model with mild tension.

It is worthwhile to note that there are two length scales $H^{-1}$ and $L_\la$ which are dynamical and non-dynamical, respectively. 
One could interpret that the behavior of the Universe is explained by comparing the aforementioned length scales. 
The RHDE behaves similarly to the scaling solution as found in the standard model of the holographic dark energy \cite{Hsu:2004ri, Li:2004rb} when $H^{-1}\ll L_\la$. 
When the Universe is sufficiently large corresponding to the magnitudes of $H^{-1}$ and $L_\la$ are in the same order, the non-extensivity due to the \Ren entropy is active, making the RHDE evolves differently. 
Eventually, the RHDE behaves similarly to the dark energy in $\Lambda$CDM model when $H^{-1}\gg L_\la$. 
In other words, the RHDE cannot make the Universe expand with acceleration if the size of the Universe is too small compared to the non-extensive length. 
After that, when the Universe evolves until its size is significantly large, i.e., $H^{-1}\gg L_\la$, the RHDE can drive the agreeing late-time evolution.

By interpreting that the RHDE behaves as a thermal system, there are conditions in which this system is thermodynamic stable as $H^{-1}/L_\la\gtrapprox1.98$. 
According to the numerical results for the evolution of the ratio $l_H/l_\la$ in the left panel of Fig.~\ref{fig:lH Ode}, the Universe in early time ($\ln a\lessapprox-2$) corresponds to the unstable phase while the Universe will undergo to the stable phase as time evolves.
From the numerical solving of the evolution of the ratio $l_H/l_\la$ with $l_{\la,\text{AIC}_\text{min}}$, the condition for the Universe being in the stable phase is $\ln a\gtrapprox -1.69$. 
The lower bound of the aforementioned period is around the end of the matter-dominated epoch.
It is possible to interpret that the Universe evolves in such a way that the thermal system undergoes from unstable to stable phases.
Moreover, from the fact that the logarithmic function in Eq.~\eqref{SR in SBH} is the monotonically increasing function, an examination of the evolution of the IR length scale (i.e. Hubble radius, for our cosmological model) shown in the left panel of Fig.~\ref{fig:lH Ode} remarkably implies an increase in the \Ren entropy over time.
The second law of \Ren black hole thermodynamics is therefore satisfied in this aspect.
As a result, this work may shed light on the interplay between the non-extensive nature of the black hole in the description of thermodynamics and the evolution of the Universe in the description of classical gravity.

Throughout our consideration, we have focused on the HDE model constructed only from the simple Schwarzschild black hole. It may be worthwhile to extend this study to other types of black holes.
With the complicated phase structures of those black holes, tantalizing implications for cosmic evolution are possibly obtained.
Additionally, the possibility of an interaction between RHDE and dark matter is also an interesting scenario for further study in order to gain insight into cosmological issues, such as fine-tuning and coincidence problems.
The aforementioned investigations are left as the future works.

As a last remark of this paper, the perturbation analysis of the holographic dark energy model can be investigated if an action for the holographic dark energy model is constructed. 
With a similar fashion as proposed in Refs.~\cite{Li:2012xf, Li:2013, Lin:2021bxv}, the consistent (mini superspace) action for our RHDE model associated with the flat FLRW ansatz: $\dd s^2=-N(t)^2\dd t^2+a(t)^2\Big(\dd r^2+r^2\dd\theta^2+r^2\sin^2\theta\dd\phi^2\Big)$ might be written down as
\ba
	S=\int\dd t\left[\sqrt{-g}\left\{\frac{R}{16\pi G}-\rho_\text{RHDE}(L(t))\right\}-\alpha(t)\left\{L(t)-\frac{a(t)}{\dot{a}(t)}\right\}\right]+S_\text{r/m},
\ea
where $R$ and $S_\text{r/m}$ are the Ricci scalar and the action for radiation/matter, respectively. 
The energy density $\rho_\text{RHDE}(L)$ is straightforwardly that from Eq.~\eqref{rho RHDE in L}. 
It will appear in the Friedmann equation~\eqref{Ein 00} obtaining from varying the action with respect to the auxiliary field $N(t)$.
Note also that the characteristic length scale $L(t)$ is constrained as the Hubble radius with a associated Lagrange multiplier $\alpha(t)$.
Based on the formulation endowed with the action for HDE model, this type of dark energy is further investigated in various aspects, e.g., perturbation analysis (see Ref.~\cite{Lin:2021bxv}).
This aforementioned analysis for RHDE is also left as another future work.


\section*{Acknowledgement}

We would like to thank Khamphee Karwan for valued discussions. This research has received funding support from the NSRF via the Program Management Unit for Human Resources \& Institutional Development, Research and Innovation [grant number B13F660066].
CP is also supported by Fundamental Fund 2566 of Khon Kaen University and Research Grant for New Scholar, Office of the Permanent Secretary, Ministry of Higher Education, Science, Research and Innovation under contract no. RGNS 64-043.


%


\end{document}